\documentclass[a4paper,12pt]{article}
\usepackage[utf8]{inputenc}
\usepackage[italian, english]{babel}
\usepackage[left=2cm,right=2cm,top=2cm,bottom=2cm]{geometry}

\usepackage{amsmath} 
\usepackage{graphicx} 
\usepackage{float}
\usepackage{amsthm} 
\usepackage{amsfonts} 
\usepackage{amssymb} 
\usepackage{xcolor} 
\usepackage{listings} 
\definecolor{backcolour}{rgb}{0.95,0.95,0.92}
\usepackage{verbatim}
\usepackage{siunitx}
\usepackage{youngtab}
\usepackage{ytableau}
\usepackage{subfig}
\usepackage{braket}
\usepackage{physics}
\usepackage{slashed}
\usepackage{booktabs}
\usepackage{tikz}
\usepackage{dynkin-diagrams}

\usepackage[T1]{fontenc} 

\usepackage{yfonts}

\numberwithin{equation}{section}

\usepackage{bbm}
\usepackage[utf8]{inputenc}

\usepackage{hyphenat}
\usepackage{wrapfig}
\usepackage{empheq}
\usepackage{textcomp}

\usepackage{rotating}

\usepackage{pdfpages}

\usepackage{setspace}
\usepackage{array}
\usepackage{caption}
\usepackage[pdf]{pstricks}
\usepackage{upgreek}
\usepackage{cmll}
\usepackage{latexsym}

\usepackage{cite}
\usepackage{epsfig}

\usepackage[titletoc,page]{appendix}
\usepackage{multicol}

\usepackage{tikz-cd}

\newcommand{\beq}{\begin{eqnarray}}
	\newcommand{\eeq}{\end{eqnarray}}
\newcommand{\bea}{\begin{eqnarray}}
	\newcommand{\eea}{\end{eqnarray}}
\newcommand{\be}{\begin{equation}}
	\newcommand{\ee}{\end{equation}}

\def\brc{\langle }
\def\ckt{\rangle}

\usepackage[colorlinks=true, linkcolor=blue, citecolor=blue]{hyperref} 

\begin{document}

\title{Chiral gauge theories, generalized anomalies and breakdown  of the 
color-flavor-locked center symmetry }

\vskip 40pt  
\author{  
Stefano Bolognesi$^{(1,2)}$, 
 Kenichi Konishi$^{(1,2)}$, Andrea Luzio$^{(3,2)}$, \\
  and  Matteo Orso$^{(1,2)}$    \\[13pt]
{\em \footnotesize
$^{(1)}$Department of Physics ``E. Fermi'', University of Pisa,}\\[-5pt]
{\em \footnotesize
Largo Pontecorvo, 3, Ed. C, 56127 Pisa, Italy}\\[2pt]
{\em \footnotesize
$^{(2)}$INFN, Sezione di Pisa,    
Largo Pontecorvo, 3, Ed. C, 56127 Pisa, Italy}\\[2pt]
{\em \footnotesize
$^{(3)}$Scuola Normale Superiore,   
Piazza dei Cavalieri, 7,  56127  Pisa, Italy}\\[2pt]
\\[1pt] 
{ \footnotesize  stefano.bolognesi@unipi.it,  kenichi.konishi@unipi.it,   andrea.luzio@sns.it,  matteo.orso@pi.infn.it }  
} 
\date{}

\vskip 6pt

\maketitle

\begin{abstract}

We study the strong-interaction dynamics of a class of $4D$ chiral $SU(N)$ gauge theories with a fermion in a symmetric second-rank tensor representation  and a number of fermions  in an anti-antisymmetric tensor representation, extending the previous work on chiral gauge theories such as the Bars-Yankielowicz and the generalized Georgi-Glashow models. The main tool of our analysis is the anomalies obstructing the gauging of certain 1-form color-flavor-locked center symmetry, together with some flavor symmetries. The matching requirement for these mixed-anomalies strongly favors dynamical Higgs phases caused by bifermion condensate formation, against  a confinement phase with no condensates
 and no symmetry breaking, or a confinement phase with multifermion color-singlet condensates only.  Dynamical gauge symmetry breaking and 
 the spontaneous breaking of a $U(1)$ symmetry caused by such condensates mean that the color-flavor-locked 1-form center symmetry itself is lost in the infrared.  One is led to a solution, if not unique,  which  
 satisfies fully the conventional as well as the new, generalized  't Hooft anomaly matching requirements.


\end{abstract}

\maketitle
	
	\newpage

	\tableofcontents

	\newpage 
		
	\section{Introduction}  
	
    The strong-interaction dynamics of chiral gauge theories in four dimensions are still not satisfactorily understood, after many years of 
 investigations \cite{Raby}-\cite{BK}.  Some hints about their infrared dynamics are however emerging from the recent studies, which take into account new kinds of 't Hooft anomalies related to the gauging of certain 1-form center symmetries  \cite{Seiberg}-\cite{Anber}. These new inputs arise from the idea of generalized symmetries, which act on extended objects such as closed loops and surfaces, as compared to the conventional symmetries transforming local operators (i.e. acting on particles).  Even though formally distinct from the conventional ones, they {\it are} symmetries of the system considered, and it is up to us to decide to  {\it gauge} them, introduce appropriate gauge functions to identify the field configurations related by the symmetries,  and eliminate the double counting.  Any topological obstruction in doing so is a 't Hooft anomaly.  As in the conventional 
't Hooft anomaly matching argument,  these new anomalies must also be appropriately reflected in the infrared.
Generally, these requirements impose stronger constraints on the way the symmetries of a given theory are realized in the infrared, than those following from the conventional 't Hooft anomaly matching argument only.  

A number of chiral gauge theories have been studied in the last several years along these lines of  thoughts.  The examples  are the Bars-Yankielowicz  models,  Georgi-Glashow type models, and some others  \cite{BKL1}-\cite{corfu}.  The results of these studies  support  the idea of  
 dynamical gauge symmetry breaking (dynamical Higgs phase, or dynamical Abelianization);   the hypothesis of confining vacua without global symmetry breaking (no condensate formation), proposed in some of these models,  based on the conventional  
perturbative 't Hooft anomaly matching,   is found to be inconsistent with the new, mixed-anomaly matching requirement.  

It is the aim of this work to extend these analyses to  some simplest 2-index tensor chiral gauge theories, where 
the matter fermions consist of some symmetric   ($\psi$)  as well as (anti-) antisymmetric  ($\chi$)  tensor representations, but without fundamentals,   namely, with fermions $\psi$ and  $\chi$  in the representation, 
 \begin{equation}
\Yvcentermath1
\frac{N-4}{\ell} \hspace{1mm} \yng(2) \oplus \frac{N+4}{\ell} \hspace{1mm} \bar{\yng(1,1)}\;,     \label{models}   
\end{equation} 
 $\ell$ being a common divisor of $N-4$ and $N+4$
($N \geq 5$).
Before the advent of the generalized symmetries and related new developments, these models have been studied by
Shi and Shrock \cite{ShiShr1, ShiShr2}, who have investigated various possible dynamical scenarios of these models, and more recently, by using the 
anomalies associated with the generalized (1-form) symmetries, briefly in \cite{BKL1} and more extensively, in \cite{AnberHong,AnberChan}.  

We shall consider here  some   $SU(N)$  2-index  tensor models (\ref{models}) with $N=5,6,8$, which become  strongly coupled in the infrared.

	\section{Color-flavor-locked 1-form  $\mathbbm Z_N $ center symmetry \label{GAnomalies}  }
		
As  the idea of color-flavor locked  ${\mathbbm Z}_N $ 1-form symmetry is central below, let us briefly review it first. 
 In general,   the color center  $\mathbbm Z_N \subset SU(N)$ symmetry is broken by the presence of fermions (unless  all the fermions present are  e.g., in the adjoint representation).      In the present paper, however,  we are interested in $SU(N)$ models with a set of massless Weyl fermions $\{\psi^k\}$ in $N$-ality 2 representations of $SU(N)$. When $N$ is even, for instance, these models enjoy a $\mathbbm Z_2 \subset \mathbbm Z_N$ center  symmetry. 
    Gauging this symmetry allows to uncover new 't Hooft anomalies, which have been exploited fruitfully in \cite{BKL1}.

  In the models under study, we can actually do better. This is possible if the action of the $SU(N)$ center on the matter fields can be canceled by some $U(1)$ symmetry transformation.  
    For this to happen, the following relation must hold: 
    \be       
    \sum_i  q_k^{(i)} \frac{m_i}{N}  +     \frac{{\cal N}_k}{N}= 0 \mod \mathbbm Z \;,    \qquad     \forall  k     \label{conditions}
    \ee
    where $q_k^{(i)}$ is the (integer) charge of the fermion $\psi^k$ under the symmetry $U(1)_i$, $\mathcal{N}_k$ is the N-ality of the $SU(N)$ representation where $\psi^k$ sits, and $m_i$ are some integer coefficients to be chosen appropriately
    ($m_i/N$ is the required $U(1)_i$ rotation angle). 
    Then, a ``color-flavor locked'' 1-form symmetry  
\cite{Tanizaki,BKL2,BKL4,BKLReview, BKLDA, BKLproc,BKLZ2,corfu, AnberHong,AnberChan} 
    acts  by shifting simultaneously  \footnote{The center symmetry is formally defined as a path-ordered sequence of local  $SU(N)$ gauge transformations along the loop:  the fermions must also transform in order to keep the action invariant.   After encircling the loop and coming back to the original point, the gauge field is transformed by a nontrivial periodicity with a $\mathbbm Z_N $ factor,  dragging  the  fermions fields to transform as in (\ref{center1}).  It would  violate their periodic boundary condition  (i.e., their uniqueness at  each spacetime point) in general, and this is the reason why the presence of fermions normally breaks the center symmetry. 
} the $SU(N)$ and the $U(1)$ holonomies: 
    \be  
    \Tr[{\cal P}  e^{i \oint_L    a }]    \to      e^{ \tfrac{2\pi  i}{N}  }    \Tr[{\cal P}  e^{i \oint_L  a }]  \;, \qquad     
       \psi^k  \to   e^{  \tfrac{ 2\pi i  {\cal N}_k  }{N}}    \psi^k   \;,    \label{center1} 
    \ee
    \be
         \, e^{i \oint_L  A_i }  \to   \left(  e^{  \frac{2\pi i}{N}  m_i }   \right) \, e^{i \oint_L  A_i } \;       \label{center2} 
    \ee 
    (here $a$ is the dynamical  $SU(N)$ gauge field, whereas $A_i$ is the background $U(1)_i$ gauge field).
    
    As it is, (\ref{center1}), (\ref{center2}),  is still a global symmetry of the model. For us, it is important that this symmetry can be gauged, i.e. we can promote (\ref{center1}), (\ref{center2}),  to an identification (of the fields related by it), at the price of introducing an external background (2-form) field, $b^{(2)}_N \in H^2(M, \mathbbm Z_N)$.    A gauge   or a $U(1)_i$   Wilson loop is not gauge invariant each;  
       however, thanks to the relation (\ref{conditions}) the full Wilson loops of the matter field connections are well-defined, gauge invariant, line operators.
    
    Geometrically, the gauging of the color-flavor-locked 1-form symmetry prescribes to path integrate on $PSU(N)$ bundles $P$ where 
    \be \omega_N(P)= b^{(2)}_N\ee
    (here $\omega_N(P)$ is a generalization of the Stiefel-Whitney class). At the same time, the gauging imposes to replace the $U(1)_i$ connection with $U(1)/\mathbbm Z_N$ connections, 
    \be
    dA_i \to dA_i  +\frac{m_i}{N} b^{(2)}_N\;.
    \ee

    In the following, we apply the well-known strategy (see \cite{KapSei,GKSW}) of implementing the $b^{(2)}_N$ field as a couple of continuous 1-form and 2-form $U(1)$ fields, $B^{(1)}$ and $B^{(2)}$ \footnote{Roughly, $b^{(2)}_{N} \sim \frac{N}{2\pi} B^{(2)}_c \,  \mod N$.},
    with the constraint
    \be
     dB^{(1)}=N B^{(2)}
    \ee
    and the continuous gauge invariance 
    \be
    B^{(1)}\to B^{(1)}+ N \lambda^{(1)}\;, \quad B^{(2)}\to B^{(2)}+ d\lambda^{(1)}\;,
    \ee
    where $\lambda^{(1)}$ is a $U(1)$ connection. 
    As the procedure is  by now well-known, we refer the reader to the original papers for more details.

    With the technology discussed here, one can tackle the case in which the flavor symmetry by itself is not simply connected. For example, there can be a global $SU(N_f)$ flavor symmetry whose center acts on the matter fields as a subgroup of a particular combination of the $U(1)$. The ideas described above can be applied with no practical modification.  

\section{Calculation of the mixed anomalies  \label{Mixed}}

       \subsection{$SU(5)$ model  \label{sec:su5}}
       
The first model is an $SU(5)$ gauge theory with matter fermions $\psi$ and $\chi$, in the representation
\begin{equation}\label{model_SU(5)}
\Yvcentermath1
\yng(2) \oplus    9  \; \bar{\yng(1,1)}\;,
\end{equation}
i.e.,   $N=5, \ell=1$ in (\ref{models}).  
The classical symmetry group is 
\begin{equation}\label{classical_flavor}
SU(9) \times U(1)_{\psi} \times U(1)_{\chi}\;,  
\end{equation}
with a
 non-anomalous combination of the two $U(1)$ groups 
\begin{equation}
U(1)_{\psi \chi}: \qquad  \psi \to e^{27i \alpha}\psi, \qquad   \chi \to e^{-7i\alpha}\chi\;,  \qquad   \alpha \in [0, 2\pi)    \label{Upc1} 
\end{equation}
There are also discrete sub-groups of  $U(1)_{\psi} \times U(1)_{\chi}$  left unbroken by the instantons, $\left( \mathbb{Z}_7 \right)_{\psi}$ and $\left( \mathbb{Z}_{27} \right)_{\chi}$, 
but both of them turn out to be part of  $U(1)_{\psi \chi}.$ 

The action of the center of the strong gauge group on the fermion  
\begin{equation}\label{center_color}
\mathbb{Z}_5: \qquad \psi \to e^{\frac{i4\pi}{5}k}\psi, \qquad \chi \to e^{-\frac{i4\pi}{5}k} \chi; \qquad k=1, \ldots, 5
\end{equation}
can be  compensated by   
\be  \mathbb{Z}_5 \subset U(1)_{\psi \chi}\;, \qquad  
  \alpha = -  \frac{2 \pi}{5}k \;;   \ee 
    this allows us to introduce the color-flavor combined  1-form  $\mathbb{Z}_5$  symmetry, as discussed  in (\ref{conditions})-(\ref{center2}). 
 The genuine symmetry of the model (\ref{model_SU(5)}) is  \footnote{The center of the 
flavor $SU(9)$ symmetry is also shared with $U(1)_{\psi \chi}$, as can be easily verified.}
\begin{equation}\label{symm_SU(5)}
SU(5) \times \frac{SU(9) \times U(1)_{\psi \chi}}{\mathbb{Z}_5 \times \mathbb{Z}_9}
\end{equation}
The matter content of this UV theory is summarized  in Table \ref{table_UV_SU(5)}.
\begin{table}[H]
\begin{center}
\begin{tabular}{cccc}
\toprule 
fields & $SU(5)$ & $SU(9)$ & $U(1)_{\psi \chi}$\\
\midrule
$\psi$ & \Yvcentermath1 $\yng(2)$ & $15 \hspace{1mm} (\cdot)$ & 27\\[1.5ex]
$\chi$ & \Yvcentermath1 $9 \hspace{1mm} \bar{\yng(1,1)}$ & \Yvcentermath1 $10 \hspace{1mm} \yng(1)$ & -7\\
\bottomrule
\end{tabular}
\caption{\footnotesize  Representations and charges of the matter content in the UV theory in respect of the continuous gauge and flavor symmetries. The multiplicities are also indicated, while $(\cdot)$ stands for the singlet representation.}\label{table_UV_SU(5)}
\end{center}
\end{table}

The   gauging of  the color-flavor locked  1-form  $\mathbbm{Z}_5$  symmetry  follows the by now well established  procedure  \cite{AhaSeiTac} - \cite{GKKS}.   

 We introduce the following gauge fields 
\begin{itemize}
\item $SU(5)$: $a_c \equiv (a_c)^a_{\mu}(x)t^adx^{\mu}$ $\mathfrak{su}(5)$-valued 1-form.
\item For $SU(9)$: $a_f \equiv (a_f)^a_{\mu}(x)t^adx^{\mu}$ $\mathfrak{su}(9)$-valued 1-form.
\item  $U(1)_{\psi \chi}$: $A_{\psi \chi} \equiv (A_{\psi \chi})_{\mu}(x)dx^{\mu}$ 1-form. 
\item   $\mathbb{Z}_5$: $B^{(1)}_c \equiv (B^{(1)}_c)_{\mu}(x)dx^{\mu}$ 1-form and
$B^{(2)}_c \equiv (B^{(2)}_c)_{\mu \nu}(x) dx^{\mu} \wedge dx^{\nu}$ 2-form.
\item For $\mathbb{Z}_9$: $B^{(1)}_f \equiv (B^{(1)}_f)_{\mu}(x)dx^{\mu}$ 1-form and
$B^{(2)}_f \equiv (B^{(2)}_f)_{\mu \nu}(x) dx^{\mu} \wedge dx^{\nu}$ 2-form.
\end{itemize}
$B^{(1)}_c$, $B^{(2)}_c$ are $U(1)$ gauge fields satisfying the constrains:
\begin{equation}\label{constraint1_col}
5B^{(2)}_c = \mathrm{d}B^{(1)}_c\;. 
\end{equation}
The 1-form gauge transformations are
\begin{equation}\label{gaugetransf1_col}
B^{(1)}_c \to B^{(1)}_c +5\lambda_c, \hspace{1cm} B^{(2)}_c \to B^{(2)}_c + \mathrm{d}\lambda_c\;,   
\end{equation}
where $\lambda_c \equiv (\lambda_c)_{\mu}(x)dx^{\mu}$ is a 1-form $U(1)$ gauge function. 
Analogously for $B^{(1)}_f$, $B^{(2)}_f$ :
\begin{equation}\label{constraint1_flav}
9B^{(2)}_f = \mathrm{d}B^{(1)}_f
\end{equation}
with gauge transformations
\begin{equation}\label{gaugetransf1_flav}
B^{(1)}_f \to B^{(1)}_f +9\lambda_f, \hspace{1cm} B^{(2)}_f \to B^{(2)}_f + \mathrm{d}\lambda_f
\end{equation}
where $\lambda_f \equiv (\lambda_f)_{\mu}(x)dx^{\mu}$ is another $U(1)$ gauge function. 

The 2-form gauge fields $B^{(2)}_c$    (and similarly $B^{(2)}_f)$)   carry non-trivial 't Hooft fluxes  
on topologically non-trivial 4-dimensional manifold such as  $\Sigma_4 \equiv \Sigma_2 \times \Sigma_2$:
\begin{equation}\label{flux_Z_5}
\frac{1}{8 \pi^2} \int_{\Sigma_4} ( B^{(2)}_c)^2 = \frac{n}{25}\;,  \quad n \in \mathbb{Z}_5 \;, \qquad 
\frac{1}{8 \pi^2} \int_{\Sigma_4} ( B^{(2)}_f)^2 = \frac{l}{81}, \quad l \in \mathbb{Z}_9
\end{equation}
To proceed, we embed $a_c$ into an $U(5)$ gauge field  such that 
\begin{equation}\label{def_U(5)}
\tilde{a}_c \equiv a_c + \frac{1}{5}B^{(1)}_c\;,\qquad   
\tilde{a}_c \to \tilde{a}_c + \lambda_c\;, 
\end{equation}
and similarly for  $a_f$ 
\begin{equation}\label{def_U(9)}
\tilde{a}_f \equiv a_f + \frac{1}{9}B^{(1)}_f\;, \qquad  
\tilde{a}_f \to \tilde{a}_f + \lambda_f\;. 
\end{equation}
The color-flavor combined 1-form  $\mathbb{Z}_5$   (and $\mathbb{Z}_9$)  symmetry   means that  the  $U(1)_{\psi \chi}$ gauge field  $A_{\psi \chi}$ simultaneously
 transforms under these gauge transformations,  as
 \begin{equation}\label{AB1}
A_{\psi \chi} \to A_{\psi \chi} -  \lambda_c   +   4 \lambda_f\;.    
\end{equation}
Now the Dirac operators for $\psi$  and  for $\chi$
\begin{equation}\label{dirac_psi1}
\mathrm{d} + \mathcal{R}_S\left( \tilde{a}_c\right) + 27 A_{\psi \chi} \;, \qquad   
\mathrm{d} + \mathcal{R}_{A^*}\left( \tilde{a}_c\right) -7 A_{\psi \chi} +\mathcal{R}_{F}\left( \tilde{a}_f\right) 
\end{equation}
where $\mathcal{R}_S$, $\mathcal{R}_{A^*}$ and $\mathcal{R}_F$  stand for symmetric, anti-antisymmetric and fundamental representation, can be  converted 
into the form invariant under the transformations, (\ref{gaugetransf1_col}),   
 by appropriately coupling the 1-form 
 gauge fields  $(B^{(1)}_c, B^{(2)}_c)$.      This leads to the invariant  gauge field tensors for   $\psi$ and $\chi$,  
 \be \mathcal{R}_S\left(F \left(  \tilde{a}_c\right) - B^{(2)}_c \right)  + 27 \left[ \mathrm{d}A_{\psi \chi} +B^{(2)}_c  -   4B^{(2)}_f\right] \;,  
 \ee
 \be   \mathcal{R}_{A^*}\left( F \left( \tilde{a}_c\right)- B^{(2)}_c \right)  +\mathcal{R}_{F}\left( F \left( \tilde{a}_f\right) - B^{(2)}_f\right)    -  7 \left[ \mathrm{d}A_{\psi \chi} +  B^{(2)}_c  - 4B^{(2)}_f\right] \;.  
 \ee
Following the  Stora-Zumino descent procedure,   
we  start by writing the $6D$ Abelian anomaly
\begin{equation}\label{6d_anomaly_SU(5)}
\begin{split}
\mathcal{A}^{6D} = \frac{2 \pi}{3! (2 \pi)^3} \int_{\Sigma_{6}} & \Tr_c\left[ \left( \mathcal{R}_S\left(F \left(  \tilde{a}_c\right) - B^{(2)}_c \right)  + 27 \left[ \mathrm{d}A_{\psi \chi} + B^{(2)}_c -   4B^{(2)}_f\right] \right)^3\right] + \\
& \begin{aligned}
 + \Tr_{c, f}\biggl[ \biggl( & \mathcal{R}_{A^*}\left( F \left( \tilde{a}_c\right)- B^{(2)}_c \right)  +\mathcal{R}_{F}\left( F \left( \tilde{a}_f\right) - B^{(2)}_f\right) +\\
& -   7 \left[ \mathrm{d}A_{\psi \chi} + B^{(2)}_c  -  4B^{(2)}_f\right] \biggr)^3\biggr]
\end{aligned}
\end{split}
\end{equation}
where $\Sigma_6$ is a 6-dimensional oriented manifold with boundary. 
We note that the terms proportional to $\Tr^c_F \left[ \left(F \left( \tilde{a}_c\right) - B^{(2)}_c \right)^3\right]$ cancel out, reflecting the anomaly-free nature of the gauge group $SU(5)$. The terms proportional to $\Tr^c_F \left[ \left(F \left(  \tilde{a}_c\right) - B^{(2)}_c \right)^2\right]\left( \mathrm{d}A_{\psi \chi} +B^{(2)}_c  -4B^{(2)}_f\right)$ cancel out, too, as  it should. 
One arrives at 
\begin{equation}\label{SU5anomaly}
\begin{split}
\frac{1}{24\pi^2}\int_{\Sigma_6} & 264375 \left( \mathrm{d}A_{\psi \chi} +B^{(2)}_c -    4B^{(2)}_f\right)^3 +  10   \Tr^f_F \left[ \left( F \left( \tilde{a}_f\right) - B^{(2)}_f\right)^3 \right] +\\
& -  210 \Tr^{f}_F \left[ \left( F \left( \tilde{a}_f\right) - B^{(2)}_f\right)^2\right]  \left( \mathrm{d}A_{\psi \chi} +  B^{(2)}_c  -  4B^{(2)}_f\right) \; .
\end{split}
\end{equation}
This expression contains all the conventional and generalized anomalies which gauging of the symmetries  (\ref{symm_SU(5)}) encounters.
  The first term comes from the  $[U(1)_{\psi \chi}]^3$ anomaly;
the second and the third terms in (\ref{SU5anomaly}) originate from the $[SU(9)]^3$ and the $[SU(9)]^2-[U(1)_{\psi \chi}]$ anomalies. respectively.

The first term of (\ref{SU5anomaly}) contains a term proportional to $\mathrm{d}A_{\psi \chi}(B^{(2)}_c)^2$.  This represents the mixed anomaly $[U(1)_{\psi \chi}]-[\mathbb{Z}_5]^2$. 
The $5D$ theory contains 
\begin{equation}\label{SU(5)_U(1)Z5_5D}
\frac{1}{8 \pi^2} 264375  \int_{\Sigma_5} A_{\psi \chi}(B^{(2)}_c)^2\;. 
\end{equation}
Considering the gauge variation
\begin{equation}\label{gauge_var_for4DSU(5)_U1}
\delta A_{\psi \chi} =  \mathrm{d} \delta A^{(0)}_{\psi \chi}\; , \qquad \delta A^{(0)}_{\psi \chi} = \alpha \; ,
\end{equation}
with $\alpha$ the parameter of the transformations (\ref{Upc1}), we arrive
 at  the  $4D$ anomaly 
\begin{equation}\label{SU(5)U1Z5}
\delta S^{4D}_{UV} = \frac{1}{8 \pi^2} 264375  \int_{\Sigma_4} \alpha (B^{(2)}_c)^2 = 10575 \, \alpha n\;;  \qquad n \in \mathbb{Z}_5 \; ,
\end{equation}
where use was made of  (\ref{flux_Z_5}).  The  $U(1)_{\psi \chi}$ becomes thus anomalous when  the 1-form $\mathbb{Z}_5$  is gauged. 
This anomaly clearly descends from the  $U(1)_{\psi \chi}^3$ anomaly.  

Similarly the gauging of the $1$-form  $\mathbb{Z}_9$ center of the $SU(9)$ flavor group, gives,   
\begin{equation}\label{SU(5)U1Z9}
\delta S^{4D}_{UV} = \frac{1}{8 \pi^2} 4230630  \int_{\Sigma_4} \alpha (B^{(2)}_f)^2 =  52230 \,  \alpha l\; ;  \qquad l \in \mathbb{Z}_9\;.
\end{equation}
There are  other types of mixed anomalies:

\begin{description}

\item  [(1)]  $[U(1)_{\psi \chi}]-[\mathbb{Z}_5]-[\mathbb{Z}_9]$, 
\begin{equation}\label{Mixed1}
\delta S^{4D}_{UV} =- \frac{2115000}{2} \frac{1}{2\pi} \int_{\Sigma_2}  B^{(2)}_c \frac{1}{2\pi} \int_{\Sigma_2} B^{(2)}_f \alpha = - 23500 \,   \alpha n_1 l_1\; ;  \qquad n_1 \in \mathbb{Z}_5\; ; \qquad l_1 \in \mathbb{Z}_9\;.
\end{equation}

\item  [(2)]  $[\left( \mathbb{Z}_7 \right)_{\psi}]-[\mathbb{Z}_5]^2$: 
\begin{equation}\label{SU(5)_4D_Z7}
\delta S^{4D}_{UV} = \frac{1}{8 \pi^2} 264375 \int_{\Sigma_4}\left( - \frac{2 \pi m}{7}\right)  (B^{(2)}_c)^2 = -10575 \, n \frac{2 \pi m}{7}\; ; \qquad n \in \mathbb{Z}_5\; ; \qquad m=1, \ldots, 7 \; ,
\end{equation}
under the variation 
\begin{equation}\label{gauge_var_for4DSU(5)_Z7}
 \delta A^{(0)}_{\psi \chi} =- \frac{2 \pi}{7}m\; ; \hspace{1cm} m=1, \ldots, 7 \; .
\end{equation}

\item  [(3)]   $[\left(\mathbb{Z}_{27} \right)_{\chi}]-[\mathbb{Z}_5]^2$:
\begin{equation}\label{SU(5)_4D_Z27}
\delta S^{4D}_{UV} = \frac{1}{8 \pi^2} 264375  \int_{\Sigma_4} \left( -\frac{8 \pi t}{27}\right)  (B^{(2)}_c)^2 = -1175 n \frac{8 \pi t}{3}\; ; \qquad n \in \mathbb{Z}_5\; ; \qquad t=1, \ldots, 27\;. 
\end{equation}
\item  [(4)]  $[\left(\mathbb{Z}_7 \right)_{\psi}]-[\mathbb{Z}_9]^2$.
\begin{equation}\label{SU(5)_4D_Z7-Z9}
\delta S^{4D}_{UV} = \frac{1}{8 \pi^2} 4230630 \int_{\Sigma_4}\left( - \frac{2 \pi m}{7}\right)  (B^{(2)}_f)^2 =-52230 \, l \frac{2 \pi m}{7}\; ; \qquad l \in \mathbb{Z}_9\; ; \qquad m=1, \ldots 7\,. 
\end{equation}

\item  [(5)]   $[\left( \mathbb{Z}_{27} \right)_{\chi}]-[\mathbb{Z}_9]^2$.
\begin{equation}\label{Mixed2}
\delta S^{4D}_{UV} = \frac{1}{8 \pi^2} 4230630  \int_{\Sigma_4}\left( - \frac{8 \pi t}{27}\right)  (B^{(2)}_f)^2 = -52230 \,  l \frac{8 \pi t}{27}\; ; \qquad l \in \mathbb{Z}_9\; ; \qquad t=1, \ldots, 27\;,
\end{equation}
\end{description}
some of which ((2) $\sim$ (5))  are simply special cases of (\ref{SU(5)U1Z5}) and   (\ref{SU(5)U1Z9}).

	\subsection{$SU(6)$ model   \label{sec:su6}}

	The second 
	model is $SU(6)$ $\psi\chi$ model with $N=6,  \ell=2$, in (\ref{models}),   that is,
	\be
	\Yvcentermath1
	\yng(2)  \oplus  5 \; \bar{\yng(1,1)}    \label{su6model}
	\ee
	The  symmetry group is 
	\be
	\frac{SU(6)_c \times SU(5)_\chi \times U(1)_{\psi\chi} \times \mathbbm Z_4}{\mathbbm Z_2 \times \mathbbm Z_3 \times \mathbbm Z_5}
	\ee
	$U(1)_{\psi\chi}$ acts as
	\be
	U(1)_{\psi\chi}: \quad \psi \to e^{5i\alpha}\psi\;, \quad \chi \to e^{-2i\alpha}\chi\;, \quad 0\le \alpha \le 2\pi\;.
	\ee
	In the denominator:
	\begin{itemize}
		\item $\mathbbm Z_2$ is the center subgroup $\mathbbm Z_2 \subset SU(6)_c$, which does not act on the fermions. 
		The gauging of this 1-form $\mathbbm Z_2$ symmetry and its consequences have been studied already  in \cite{BKL1}.  
		\item $\mathbbm Z_3$ is the $\mathbbm Z_3 \subset SU(6)_c$, $ e^{\frac{2\pi i}{3} k} \cdot \mathbbm 1_{6}$, $k=1, 2, 3$, whose action  on the fermions can be undone by $U(1)_{\psi\chi}$.
		\item $\mathbbm Z_5$ is the $SU(5)$ center, $e^{\frac{2\pi i}{5} k} \cdot \mathbbm 1_{5}$, $k=1,\ldots,5$, whose action on the fermions  is undone by $e^{3 i \frac{2\pi}{5} k}\in U(1)_{\psi\chi}$.
	\end{itemize}
	In the numerator, $\mathbbm Z_4$ acts as
	\be
	\mathbbm Z_4: \quad \psi\to e^{\frac{2\pi i}{4} k}\psi\;, \quad  \chi \to e^{-\frac{2\pi i}{4} k}\chi \; , \quad k=1, \ldots, 4\;. \label{eq:Z4}
	\ee
	
	It is convenient to parametrize the entire torus $U(1)_\psi \times U(1)_\chi$, in terms of the two symmetries, $U(1)_{\psi\chi}$ and $U(1)_a$, 
	\be
	U(1)_a: \qquad \psi \to e^{-7i \beta} \psi\; , \qquad \chi \to e^{3i\beta}\chi\;, \quad \beta \in (0, 2\pi)\;.
	\ee
	The two symmetries wrap the torus only once, i.e. they do not overlap. A quick way to check it is to notice that the charge matrix is unimodular,  \be
	\det \left[\begin{array}{cc}5 & -2\\ -7 & 3 \end{array}\right]=1\;.
	\ee
	The ABJ anomaly leave $U(1)_{\psi\chi}$ exact, but breaks $U(1)_{a}\to \mathbbm Z_4$, generated by 
	\be
	\psi \to e^{-7\frac{2\pi i}{4}k}\psi=e^{\frac{2\pi i}{4}k}\psi\;, \qquad \chi \to e^{3\frac{2\pi i}{4}k}\chi=e^{-\frac{2\pi}{4}k}\chi\;, \quad k=1, \ldots, 4\;. \label{Ua1}
	\ee
	exactly as in \ref{eq:Z4}. As $U(1)_{\psi\eta}$ do not overlap with $U(1)_a$, by construction, we know that it does not overlap with $\mathbbm Z_4$ either: the ABJ free subgroup of $U(1)_\psi\times U(1)_\eta$ is $U(1)_{\psi\chi}\times \mathbbm Z_4$.
	
	It is useful to compute the anomalies using this embedding $\mathbbm Z_4 \hookrightarrow U(1)_a$. 
	
\subsubsection{Gauging the $1$-form symmetries}

Following the standard procedure, we introduce the three 2-form gauge fields:
	\be
	2B^{(2)}_2=dB^{(1)}_2\;, \quad 3B^{(2)}_3=dB^{(1)}_3\;, \quad 5B^{(2)}_f=dB^{(1)}_f\;,
	\ee
The $SU(6)$ and the $SU(5)$ connections are accordingly promoted to $U(6)$ and $U(5)$ connections, $\tilde a_c$ and $\tilde a_f$,  
	\be
	\tilde a_c=a_c+ \frac{1}{2} B^{(1)}_2 + \frac{1}{3} B^{(1)}_3\;, \qquad \tilde a_f=a_f+ \frac{1}{5} B^{(1)}_f\;.
	\ee	
The system is required to be invariant under the  $1$-form  gauge transformations,  
	\be
	B^{(2)}_2 \to B^{(2)}_2+d\lambda_2\;, \quad B^{(2)}_3 \to B^{(2)}_3+d\lambda_3\;, \quad B^{(2)}_f \to B^{(2)}_f+d\lambda_f\;, 
	\ee	
	\be   \tilde a_c \to \tilde a_c + \lambda_2 + \lambda_3\;,   \qquad \tilde a_f \to \tilde a_f + \lambda_f  \;,   \ee
	and 
       \be  	A_{\psi\chi} \to A_{\psi\chi} - \lambda_3 + 3\lambda_f\;.\label{AB2}
	\ee
	The covariant derivatives for the different fields read, 
	\bea
	\psi\; :\quad    &&  d + \left(\tilde a-\frac{1}{2} B_2^{(1)} - \frac{1}{3} B_3^{(1)}\right)_{S} + 5\left(A_{\psi\chi} + \frac{1}{3}B^{(1)}_3 -\frac{3}{5} B_f^{(1)}\right) - 7A_a\;,\\
	\chi\;: \quad   &&   d + \left(\tilde a-\frac{1}{2} B_2^{(1)} - \frac{1}{3} B_3^{(1)}\right)_{\bar{A}} 
	+  \left(    {\tilde a}_f -   \frac{1}{5}	B_f^{(1)}      \right)_F 
	\nonumber    \\   
	&&- 2 \left(A_{\psi\chi} + \frac{1}{3}B^{(1)}_3 -\frac{3}{5} B_f^{(1)}\right) + 3 A_a\;.
	\eea
	The invariant tensors are
	\be
	F \left( \tilde{a}_c\right) -  B^{(2)}_2 -B^{(2)}_3\;, \quad \tilde F(\tilde{a}_f) -  B^{(2)}_f\;, \quad dA_{\psi\chi} +   B^{(2)}_3 - 3 B^{(2)}_f \;.
	\ee
The  $6D$ UV-anomaly functional is  then 
\bea
\mathcal{A}^{6D} &=& \frac{2 \pi}{3! (2 \pi)^3} \int_{\Sigma_{6}}   \Tr_c \biggl[  \left( \mathcal{R}_S\left(F \left( \tilde{a}_c\right) - B^{(2)}_2     -B^{(2)}_3\right)  + 5 \left(dA_{\psi\chi} +   B^{(2)}_3 - 3 B^{(2)}_f \right) - 7  \, d A_a \right)^3 \biggr]    \nonumber \\ 
&+& 
  \Tr_{c, f} \biggl[  \biggl( \mathcal{R}_{A^*}\left( F \left( \tilde{a}_c\right)- B^{(2)}_2 -B^{(2)}_3 \right) + \mathcal{R}_{F} \left( F\left( \tilde{a}_f \right)-B^{(2)}_f \right)       \nonumber \\    
& - &  2 \left( dA_{\psi\chi} +   B^{(2)}_3 - 3 B^{(2)}_f \right) + 3  \,  d A_a \biggr)^3
\biggr]
\eea  
where $\Sigma_6$ is a 6-dimensional oriented manifold with boundary.

		After expanding and reorganizing terms one has
	\bea
	\mathcal{A}^{UV}= \frac{1}{24\pi^2}\left\{ 3 \, \kappa_1 \Tr_c    \left[ \mathcal{R}_S \left(F \left( \tilde{a}_c\right) - B^{(2)}_2     -B^{(2)}_3 \right)^2\right] dA_a \right.     \nonumber \\
	+\kappa_2  \,\mathcal{R}_{F}   \left[\left( F\left( \tilde{a}_f \right)-B^{(2)}_f \right)  ^3\right] +    \nonumber \\
	+\, \kappa_3 \,  \mathcal{R}_{F}  \left[\left( F\left( \tilde{a}_f \right)-B^{(2)}_f \right)  ^2\right](dA_{\psi\chi} +   B^{(2)}_3 - 3 B^{(2)}_f) 
	+    \nonumber    \\
	+3\,\kappa_4 \, \mathcal{R}_{F}  \left[\left( F\left( \tilde{a}_f \right)-B^{(2)}_f \right)  ^2\right]dA_a+    \nonumber    \\
	+\kappa_5  \left(dA_{\psi\chi} +   B^{(2)}_3 - 3 B^{(2)}_f\right)^3+    \nonumber   \\
	+3\,\kappa_6  \left(dA_{\psi\chi} +   B^{(2)}_3 - 3 B^{(2)}_f\right)^2 dA_a+       \label{U(1)^3}   \\
	+3 \, \kappa_7  \left(dA_{\psi\chi} +   B^{(2)}_3 - 3 B^{(2)}_f\right) (dA_a)^2+     \nonumber    \\
	+\left. \phantom{\kappa_1 tr\left[\left(\tilde f_c -  B^{(2)}_2 -B^{(2)}_3\right)^2\right] dA}    \kappa_8  (dA_a)^3\right\}  \nonumber  
	  \eea
	where $\kappa_1, \ldots, \kappa_8$ are the usual triangular anomaly coefficients, 
	\bea
	&&   \kappa_1=4\;,\;\; \kappa_2= 15\;,\;\; \kappa_3= -30\;,\;\; \kappa_4= 45\;,\;\; \kappa_5= 2025\;,\;\; \nonumber \\  
	&&    \kappa_6= -2775\;,\;\; \kappa_7= 3795\;,\;\; \kappa_8=-5178\;.
	\eea
The $6D$ anomaly functional (\ref{U(1)^3}) contains all the information about the conventional and new, generalized anomalies of the 
$SU(6)$ theory under consideration.

The most significant of the anomalies (\ref{U(1)^3}) concerns the 
 $U_{\psi\chi}(1) -  B^{(2)}_3 $ anomalies. 
	Indeed,    (\ref{U(1)^3})     leads to a  $4D$  anomaly
	\be    45\cdot   \frac{1}{8\pi^2}  \int_{\Sigma_4}    ( B^{(2)}_3)^2  \cdot  \delta A_{\psi\chi}    =   5 \cdot  \delta A_{\psi\chi} \;. 
	\ee
	$U_{\psi\chi}(1)$   is broken   by  the gauging of the  1-form ${\mathbbm Z}_3 \subset {\mathbbm Z}_6 \subset SU(6).$
	This is analogous to  (4.22)  of  \cite{BKLDA}.  
	
	This result is of  particular  interest   in view of the result of 
	  \cite{BKL1},   where it was noted that  
	gauging of the 1-form ${\mathbbm Z}_2 \subset {\mathbbm Z}_6$ symmetry which did not act on the fermions, 
	did not imply the breaking of  $U_{\psi\chi}(1)$ symmetry.   The latter becomes anomalous only after the  {\it color-flavor-locked}   
	1-form ${\mathbbm Z}_3 \subset {\mathbbm Z}_6$ symmetry is gauged, as done here.

 The $\mathbbm Z_4$ symmetry (\ref{eq:Z4})  suffers from certain anomalous variations. 
	By collecting the terms $dA_a B^{(2)}_i B^{(2)}_j$, where $i, j$ are $2, 3, 5$, and  by considering appropriate  $U_a(1)$ transformations  (\ref{Ua1}) with $\alpha=2\pi/4 k$,  one finds	 
	\begin{itemize}
		\item[(1)]       $\mathbbm Z_4-\mathbbm Z_2^2$ 
		\be   \label{Mixed3}
		\Delta \Gamma=-6 \cdot \frac{2\pi}{4} \cdot k 
		\ee
		This is trivial only for $k$ even, signaling that, on the toron background, $\mathbbm Z_4 \to \mathbbm Z_2$ (our earlier result \cite{BKL1}).
		
		\item[(2)]  $\mathbbm Z_4-\mathbbm Z_3^2$
		\be
		\Delta \Gamma=-311 \cdot \frac{2\pi}{4} \cdot k 
		\ee
		This, if correct, signals that $\mathbbm Z_4$ is completely broken when the $\mathbbm Z_3$ CFL flux is switched on.

		\item[(3)]   $\mathbbm Z_4-\mathbbm Z_5^2$
				\be
		\Delta \Gamma=-252 \cdot {2\pi} \cdot k \;. 
		\ee
		This is trivial, meaning that $\mathbbm Z_4 $  is not affected by the 1-form gauging of the center  $\mathbbm Z_5$.

		\item[(4)]  $\mathbbm Z_4-\mathbbm Z_2 -\mathbbm Z_3$
		\be
		\Delta \Gamma=-  2  \cdot   {2\pi} \cdot k \;.
		\ee
		This is again  trivial for any $k$,  therefore this background does not break $\mathbbm Z_4$.
		
		\item[(5)]   $\mathbbm Z_4-\mathbbm Z_3 -\mathbbm Z_5$
		\be
		\Delta \Gamma=555 \cdot \frac{2\pi}{2} \cdot k \;:    \label{Mixed4}
		\ee
		 $\mathbbm Z_4$  is broken to  $\mathbbm Z_2$.

%
%
%
	\end{itemize}

		\subsection{$SU(8)$  model   \label{sec:su8}}
	
	The matter fermions in this model ($N=8,  \ell=4$, in (\ref{models}))    is 
	\begin{equation}\label{model_SU(8)_4}
\Yvcentermath1
\yng(2) \oplus 3 \; \bar{\yng(1,1)}\;.
\end{equation}
The classical symmetry is
\begin{equation}\label{classical_flavor_8-4}
SU(3) \times U(1)_{\psi} \times U(1)_{\chi}\;. 
\end{equation}
A non-anomalous $U(1)$  symmetry is 
\begin{equation}\label{U(1)_non_an_8-4}
U(1)_{\psi \chi}: \qquad \psi \to e^{9i \alpha}\psi, \qquad \chi \to e^{-5i\alpha}\chi; \qquad \alpha \in [0, 2\pi)\;. 
\end{equation}
There are also other sub-groups left unbroken by the instantons
\begin{equation}\label{Z_psi_8-4}
\left( \mathbb{Z}_{10} \right)_{\psi}: \qquad \psi \to e^{\frac{i2\pi}{10}m}\psi\; , \qquad \chi \to \chi\; ; \qquad m=1, \ldots, 10\;,
\end{equation}
\begin{equation}\label{Z_chi_8-4}
\left( \mathbb{Z}_{18} \right)_{\chi}: \qquad \psi \to \psi\; , \qquad \chi \to e^{\frac{i2\pi}{18}t} \chi\; ; \qquad t=1, \ldots, 18\,.
\end{equation}
 The symmetry is found to be  (in accordance with \cite{BKL1})
\begin{equation}\label{genuine_flavor_8-4}
SU(3) \times U(1)_{\psi \chi} \times \mathbb{Z}_2\;.
\end{equation}
Following \cite{BKL1},  $\mathbb{Z}_2$  can be chosen as:
\begin{equation}\label{Z_2_on_the_fermios_8-4}
\mathbb{Z}_2: \qquad \psi \to -\psi\; , \qquad \chi \to \chi \; .
\end{equation}
The centers are $\mathbb{Z}_8 \subset SU(8)$ and $\mathbb{Z}_3 \subset SU(3)$, 
\begin{equation}\label{center_color_8-4}
\mathbb{Z}_8: \qquad \psi \to e^{\frac{i4\pi}{8}k}\psi\; , \qquad \chi \to e^{-\frac{i4\pi}{8}k} \chi\; ; \qquad k=1, \ldots, 8 \; .
\end{equation}
\begin{equation}\label{center_flavor_8-4}
\mathbb{Z}_3: \qquad \psi \to \psi\; , \qquad \chi \to e^{\frac{i2\pi}{3}p} \chi\; ; \qquad p=1, 2, 3\,.  
\end{equation}
These center transformations on the fermions are reproduced by  $U(1)_{\psi \chi}$ transformations
\begin{equation}\label{reproducing_Z8_8-4}
\alpha = \frac{\pi}{2}k\; ; \hspace{1cm}  k=1, \ldots, 8\;, 
\end{equation}
and 
\begin{equation}\label{reproducing_Z3_8-4}
\alpha = \frac{2 \pi}{3}p\; ; \hspace{1cm}  p=1, 2, 3\;, 
\end{equation}
respectively,  so that 
the correct symmetry of the model  is 
\begin{equation}\label{symm_SU(8)_4}
SU(8) \times \frac{SU(3) \times U(1)_{\psi \chi} \times \mathbb{Z}_2}{\mathbb{Z}_8 \times \mathbb{Z}_3}\;. 
\end{equation}

The gauge fields introduced are: 
\begin{itemize}
\item For $SU(8)$: $a_c \equiv (a_c)^a_{\mu}(x)t^adx^{\mu}$ $\mathfrak{su}(8)$-valued 1-form.
\item For $SU(3)$: $a_f \equiv (a_f)^a_{\mu}(x)t^adx^{\mu}$ $\mathfrak{su}(3)$-valued 1-form.
\item For $U(1)_{\psi \chi}$: $A_{\psi \chi} \equiv (A_{\psi \chi})_{\mu}(x)dx^{\mu}$ 1-form.
\item For $\mathbb{Z}_2$: $A_2 \equiv (A_2)_{\mu}(x)dx^{\mu}$ 1-form.
\item For $\mathbb{Z}_8$: $B^{(1)}_c \equiv (B^{(1)}_c)_{\mu}(x)dx^{\mu}$ 1-form and
$B^{(2)}_c \equiv (B^{(2)}_c)_{\mu \nu}(x) dx^{\mu} \wedge dx^{\nu}$ 2-form.
\item For $\mathbb{Z}_3$: $B^{(1)}_f \equiv (B^{(1)}_f)_{\mu}(x)dx^{\mu}$ 1-form and
$B^{(2)}_f \equiv (B^{(2)}_f)_{\mu \nu}(x) dx^{\mu} \wedge dx^{\nu}$ 2-form.
\end{itemize}
$B^{(1)}_c$, $B^{(2)}_c$ and $B^{(1)}_f$, $B^{(2)}_f$  are all $U(1)$ gauge fields which satisfy the following constrains:
\begin{equation}\label{constraint1_8-4}
8B^{(2)}_c = \mathrm{d}B^{(1)}_c\;, \qquad  
3B^{(2)}_f = \mathrm{d}B^{(1)}_f \; ,
\end{equation}
where the 1-form center transformations act as 
\begin{equation}\label{gaugetransf1_8-4}
B^{(1)}_c \to B^{(1)}_c +8\lambda_c\; , \hspace{1cm} B^{(2)}_c \to B^{(2)}_c + \mathrm{d}\lambda_c \; ,
\end{equation}
\begin{equation}\label{gaugetransf_forZ_3_8-4}
B^{(1)}_f \to B^{(1)}_f +3\lambda_f\; , \hspace{1cm} B^{(2)}_f \to B^{(2)}_f + \mathrm{d}\lambda_f \; .
\end{equation}
The 2-forms gauge fields $B^{(2)}_c$ and $B^{(2)}_f$ carry respectively a non-trivial 't Hooft flux:
\begin{equation}\label{flux_Z8_2D_8-4}
\frac{1}{2 \pi} \int_{\Sigma_2} B^{(2)}_c = \frac{n_1}{8}\;,  \quad n_1 \in \mathbb{Z}_8\;; \qquad 
\frac{1}{2 \pi} \int_{\Sigma_2} B^{(2)}_f = \frac{l_1}{3}\;,  \quad l_1 \in \mathbb{Z}_3 \; ,
\end{equation}
where $\Sigma_2$ is a closed 2-dimensional manifold. On topologically non-trivial 4-dimensional manifold $\Sigma_4 \equiv \Sigma_2 \times \Sigma_2$, they carry
\begin{equation}\label{flux_Z_8_8-4}
\frac{1}{8 \pi^2} \int_{\Sigma_4} ( B^{(2)}_c) ^2 = \frac{n}{64}\;, \quad n \in \mathbb{Z}_8\;;  \qquad 
\frac{1}{8 \pi^2} \int_{\Sigma_4} ( B^{(2)}_f) ^2 = \frac{l}{9}\; ; \qquad l \in \mathbb{Z}_3 \; .
\end{equation}
$a_c$ is  embedded in  $U(8)$
\begin{equation}\label{def_U(8)_4}
\tilde{a}_c \equiv a_c + \frac{1}{8}B^{(1)}_c \;;   \qquad 
\tilde{a}_c \to \tilde{a}_c + \lambda_c \;,
\end{equation}
and similarly  
\begin{equation}\label{def_U(3)_flavor}
\tilde{a}_f \equiv a_f + \frac{1}{3}B^{(1)}_f \;;  \qquad 
\tilde{a}_f \to \tilde{a}_f + \lambda_f\;.
\end{equation}
$A_{\psi \chi}$ transforms as
\begin{equation}\label{AB3}
A_{\psi \chi} \to A_{\psi \chi} - 2\lambda_c -\lambda_f\;.   
\end{equation}
The Dirac operator for $\psi$ is  
\begin{equation}\label{dirac_psi1_8-4}
\mathrm{d} + \mathcal{R}_S\left( \tilde{a}_c\right) + 9 A_{\psi \chi}  + A_2\; .
\end{equation}
Its invariant form is  
\begin{equation}\label{dirac_psi2_8-4}
\mathrm{d} + \mathcal{R}_S\left( \tilde{a}_c\right) + 9 A_{\psi \chi} + A_2 +2  B^{(1)}_c  +3B^{(1)}_f  \; .
\end{equation}
Similarly,  
\begin{equation}\label{dirac_chi2_8-4}
\mathrm{d} + \mathcal{R}_{A^*}\left( \tilde{a}_c\right) -5 A_{\psi \chi} +\mathcal{R}_{F}\left( \tilde{a}_f\right)  -B^{(1)}_c - 2 B^{(1)}_f 
\end{equation}
for each $\chi$  field.

The gauge field strength is 
\begin{equation}\label{F_psi_SU(8)-4}
\begin{split}
& \mathcal{R}_S\left(F \left(  \tilde{a}_c\right)\right)  + 9 \, \mathrm{d}A_{\psi \chi} +\mathrm{d}A_2  +16  B^{(2)}_c +3\cdot 3B^{(2)}_f = \\ 
& = \mathcal{R}_S\left(F \left(  \tilde{a}_c\right) - B^{(2)}_c  \right)  + 9 \left[ \mathrm{d}A_{\psi \chi} +2B^{(2)}_c +B^{(2)}_f  \right] +\mathrm{d}A_2
\end{split}
\end{equation}
for $\psi$, and 
\begin{equation}\label{F_chi_SU(8)-4}
 \mathcal{R}_{A^*}\left( F \left( \tilde{a}_c\right)- B^{(2)}_c  \right)  +\mathcal{R}_{F}\left( F \left( \tilde{a}_f \right) - B^{(2)}_f  \right) +5 \left[ -\mathrm{d}A_{\psi \chi} -2B^{(2)}_c -B^{(2)}_f \right] 
\end{equation}
for  $\chi$. 
The $6D$ Abelian anomaly functional is
\begin{equation}\label{6d_anomaly_SU(8)-4}
\begin{split}
\mathcal{A}^{6D} = \frac{2 \pi}{3! (2 \pi)^3} \int_{\Sigma_{6}} & \Tr_c \biggl[  \left( \mathcal{R}_S\left(F \left( \tilde{a}_c\right) - B^{(2)}_c  \right)  + 9 \left( \mathrm{d}A_{\psi \chi} +2B^{(2)}_c +B^{(2)}_f \right) + \mathrm{d}A_2 \right)^3 \biggr] +\\ 
& \begin{aligned}
 + \Tr_{c, f} \biggl[  \biggl( &  \mathcal{R}_{A^*}\left( F \left( \tilde{a}_c\right)- B^{(2)}_c \right) + \mathcal{R}_{F}\left( F \left( \tilde{a}_f \right)-B^{(2)}_f \right)+\\
& + 5 \left(- \mathrm{d}A_{\psi \chi} -2B^{(2)}_c -B^{(2)}_f \right)  \biggr)^3
\biggr] \; ,
\end{aligned}
\end{split}
\end{equation}
where $\Sigma_6$ is a 6-dimensional oriented manifold with boundary. 
The final expression of the $6D$ anomaly is 
\begin{equation}\label{count_3_anomaly_8-4}
\begin{split}
\frac{1}{24\pi^2} \int_{\Sigma_{6}} &15744 \left( \mathrm{d}A_{\psi \chi}  +2B^{(2)}_c+B^{(2)}_f  \right)^3 +  36(\mathrm{d}A_2)^3 +
 30 \Tr^c_F \left[ \left(F \left(  \tilde{a}_c\right) - B^{(2)}_c  \right)^2 \right] \mathrm{d}A_2 +\\
&+ 8748 \left( \mathrm{d}A_{\psi \chi} +2B^{(2)}_c + B^{(2)}_f \right)^2\mathrm{d}A_2 
+972  \left( \mathrm{d}A_{\psi \chi}  +2B^{(2)}_c +B^{(2)}_f \right)(\mathrm{d}A_2)^2  +\\
&+  28\, \Tr^{f}_F \left[ \left( F \left( \tilde{a}_f \right)-B^{(2)}_f \right)^3 \right] 
+420 \, \Tr^{f}_F \left[ \left( F \left( \tilde{a}_f \right) - B^{(2)}_f \right)^2 \right] \left( -\mathrm{d}A_{\psi \chi}  -2B^{(2)}_c -B^{(2)}_f \right)
\end{split}
\end{equation}
which contains all the information about the conventional as well as the generalized anomalies.  

The most significant anomaly  comes from the mixed term,  
  $[U(1)_{\psi \chi}]-[\mathbb{Z}_8]^2$.     The $5D$ theory contains the piece
\begin{equation}\label{5D_U(1)-Z8_8-4}
\frac{1}{8\pi^2} 62976 \int_{\Sigma_5}A_{\psi \chi}(B^{(2)}_c)^2 \; .
\end{equation}
The gauge variation is
\begin{equation}\label{gauge_variation_forU(1)_8-4}
\delta A_{\psi \chi} =  \mathrm{d} \delta A^{(0)}_{\psi \chi}\; , \hspace{1cm} \delta A^{(0)}_{\psi \chi} = \alpha \; ,
\end{equation}
with $\alpha$ the parameter of the transformation (\ref{U(1)_non_an_8-4}). By anomaly inflow, the variation of the $4D$ UV action follows 
\begin{equation}\label{4D_U(1)-Z8_8-4}
\delta S^{4D}_{UV} = \frac{1}{8\pi^2} 62976 \int_{\Sigma_4}\alpha (B^{(2)}_c)^2 = 984 \alpha n\; ;  \qquad n \in \mathbb{Z}_8\;. 
\end{equation}
 $U(1)_{\psi \chi}$ symmetry  is broken by the  gauging  of the 1-form  $\mathbb{Z}_8$ center of the strong gauge group $SU(8)$.

Analogously, the gauging of the  flavor  $SU(3)$  center leads to 
\begin{equation}\label{4D_U(1)-Z3_8-4}
\delta S^{4D}_{UV} = \frac{1}{8\pi^2} 16164 \int_{\Sigma_4}\alpha (B^{(2)}_f)^2 =1796\,  \alpha l\; ;  \qquad l \in \mathbb{Z}_3\;.
\end{equation}

Other mixed  anomalies include
\begin{description}

\item  [(1)]   $[\mathbb{Z}_2]- [\mathbb{Z}_8]^2$:
\begin{equation}\label{Mixed5}
\delta S^{4D}_{UV} = \frac{1}{8\pi^2} 11584 \int_{\Sigma_4} (B^{(2)}_c)^2 (\pm \pi) = 181(\pm \pi)n\; ; \qquad n \in \mathbb{Z}_8\;;
\end{equation}

\item  [(2)]   $[\mathbb{Z}_2]-[\mathbb{Z}_3]^2$ :
\begin{equation}\label{4D_Z3-Z2_8-4}
\delta S^{4D}_{UV} = \frac{1}{8\pi^2} 2916 \int_{\Sigma_4}(B^{(2)}_f)^2(\pm \pi) = 324(\pm \pi)l\; ; \qquad l \in \mathbb{Z}_3\;; 
\end{equation}

\item [(3)] $[\mathbb{Z}_8]-[\mathbb{Z}_3]-[\mathbb{Z}_2]$:
\begin{equation}\label{Mixed6}
\delta S^{4D}_{UV} = \frac{11664}{2} \frac{1}{2\pi}\int_{\Sigma_2}B^{(2)}_c\frac{1}{2\pi}\int_{\Sigma_2}B^{(2)}_f (\pm \pi) = 243(\pm \pi)n_1l_1\; ; \qquad n_1 \in \mathbb{Z}_8\; ; \quad l_1 \in \mathbb{Z}_3\;. 
\end{equation}

\end{description}

\section{The breakdown of the color-flavor-locked  $1$-form ${\mathbbm Z}_N$  center symmetries \label{sec:implication}}    

Our task now is to understand what the various mixed anomalies calculated above tell us about the infrared  physics of these models.  
It is essential to keep in mind that whatever nontrivial physics occurs in the infrared (the phase and symmetry of the system) it is caused by the strong-interaction dynamics of $G_c$ interactions.
 Therefore the principal question is which dynamical possibility is the most likely one, consistent with  the results found in Sec.~\ref{Mixed}. 
Understanding this, in a sense, transcends the more general, otherwise reasonable-sounding question: how are those mixed anomalies calculated in Sec.~\ref{sec:su5}, Sec.~\ref{sec:su6}, Sec.~\ref{sec:su8} realized (or matched)  in the infrared?     Let us explain this. 

These anomalies  have been recently calculated also by Anber and Chan \cite{AnberChan},  for {\it all}   asymptot\-ically-free (AF)  2-index tensor chiral theories,  (\ref{models}),    of which the  $SU(5), SU(6), SU(8)$ models studied in this note are among the simplest, and among the most strongly coupled,  systems.  \cite{AnberChan} formulates the problem as that of finding the solution of the generalized cocycle conditions, allowing for all possible weak gauging of the flavor (as well as color)  symmetries. Though formally it looks distinct,  the content of those conditions is equivalent to the 1-form gauge invariance we
imposed explicitly in the UV actions.  Our results, whenever the comparison is possible, indeed agree with those found in \cite{AnberChan}.  

The authors of   \cite{AnberChan}  assume that the system,   if one excludes possible models which might flow into a conformal (CF) fixed point,  
\begin{description}
  \item[(1)]   either   confines, with no condensate formation, and all flavor symmetries surviving in the infrared,  
     or  
  \item[(2)]  confines, with multifermion condensates, {\it all  $G_c$-gauge invariant},  which break the flavor symmetry in various ways, depending on the condensates which can be formed.  
  \end{description}
In both cases,  the conventional as well as the new, generalized 't Hooft anomaly matching conditions must be satisfied, with respect to all unbroken flavor symmetries, by the fermions present in the IR theory. By assumption, the latter are gauge-invariant composite fermions, ``baryons''.   The problem is that in most of the models analyzed  (models with $N \ge 8$, and called bosonic theories in \cite{AnberChan}) gauge-invariant multifermion operators are all bosonic, thus the option (1) is simply out of question. 
Even for other models, no appropriate set of baryons can be found, so option (1) is excluded anyway. 
The conclusion of  \cite{AnberChan} is that it is extremely difficult, if not impossible, even with the assumption (2),   to find a reasonable solution.   For instance, for $SU(5), \ell=1$ model, a multifermion condensate  (made of 12 fermions) 
is proposed, which would lead to an IR system with an anomaly-free symmetry group.

~~~

Our idea how the strong $G_c$ dynamics determines the IR physics in these models,   
differs from that of  \cite{AnberChan}.
 Excluding a priori gauge noninvariant condensates  would be to deny the existence of  the Higgs phase 
  in general \footnote{Usually, Higgs mechanism is taught in the context of a weakly coupled gauge systems, and with a Higgs scalar field 
included in the Lagrangian, with an appropriate potential. But the essence of the Higgs mechanism is the same even if the scalar Higgs field appears as a (e.g.,) bifermion composite. Even in QCD, at high densities,   the vacuum is believed to be in  color-superconducting phase, with bifermion condenstes,  $ \brc \psi_L  \psi_L  \ckt   \ne 0\;,  \,    \brc \psi_R  \psi_R  \ckt \ne 0$\;.
 }. 

 We believe that the most significant message from the  mixed anomalies calculated  in Sec.~\ref{GAnomalies}  
is that  the $U_{\psi\chi}(1)$   symmetry,  which is nonanomalous under the conventional quantization (the instantons),  becomes anomalous \footnote{Exactly the same occurs in the so-called $\psi\chi\eta$ model \cite{BKLDA}. In certain Bars-Yankielovicz and generalized Georgi-Glashow models, the same role as $U_{\psi\chi}(1)$ here  is played by a ${\mathbbm Z}_2$ symmetry  \cite{BKL2,BKL4,BKLZ2}.   
    }.    And this can be interpreted as   an indication that  {\it        $U_{\psi\chi}(1)$ symmetry is spontaneously broken by the strong  $G_c$  interactions.}   In other words, our proposal  is that 
\begin{description}
  \item[(3)]  certain bifermion condensate  such as  $\brc \psi \chi  \ckt $  or   $\brc \chi \chi  \ckt$   forms,  and breaks dynamically (part or all of) the color $G_c$ and flavor  symmetries, including  $U_{\psi\chi}(1)$.
\end{description}
Once this is assumed,  one is led naturally to the IR  system  where the massless fermions present 
satisfy {\it   all} the  anomalies, conventional as well as the generalized, mixed and global, with respect to the symmetries which are not  spontaneously broken. How this  (the Natural Anomaly Matching \cite{BKLO}) works  is reviewed below  in 
  Section \ref{sec:NAM}, and will be further illustrated in concrete examples in Sec.~\ref{DSB}.  
  
  ~~~
 
The dynamical color gauge symmetry breaking on the one hand, and the spontaneous breaking of the  $U_{\psi\chi}(1)$  symmetry on the other, 
mean that  {\it   the color-flavor locked  $1$-form center symmetries  (${\mathbbm Z}_5$ in the $SU(5)$ model,  ${\mathbbm Z}_3$ in the $SU(6)$ model, ${\mathbbm Z}_8 \subset SU(8)$) are also lost.}   Let us recall  that  the color-flavor locked  $1$-form center symmetries involve both the color ${\mathbbm Z}_N \subset SU(N)$ center transformation and a  $U_{\psi\chi}(1)$  gauge transformation  simultaneously,  see (\ref{center1}), (\ref{center2}).     
 The dynamical color gauge symmetry breaking means that   the system is brought to a Higgs (unconfined) phase.
At the same time,  the global $U_{\psi\chi}(1)$ symmetry is spontaneously broken by the bifermion condensate, and so is the $U_{\psi\chi}(1)$ gauge symmetry  which should accompany
 the color  ${\mathbbm Z}_N$ transformation.
  The color-flavor locked  $1$-form center symmetries do not survive in the infrared.

The breakdown  of the  color-flavor locked   $1$-form center symmetries  in the infrared  
   means that the details of all different sorts of mixed anomalies involving 
various discrete symmetries enlisted  in Sec.~\ref{sec:su5} $\sim$ Sec.~\ref{sec:su8} (also in \cite{AnberChan}),  are  actually immaterial  for the determination of the IR  properties of the system \footnote{  
 For instance, the presence or absence of  mixed anomalies involving $\left( \mathbb{Z}_7    \right)_{\psi}$  or  $\left(\mathbb{Z}_{27} \right)_{\chi}$ in the $SU(5)$ model,    (\ref{Mixed1}) - (\ref{Mixed2}),   those containing ${\mathbbm Z}_4$ in the $SU(6)$ model,  (\ref{Mixed3}) - (\ref{Mixed4}), and so on,  does not imply that these discrete symmetries are  broken (or respected)  by the bifermion condensates.}. 
What the mixed anomalies tell us is that it is not possible to gauge these discrete symmetries, {\it  together with}  the $1$-form color (or flavor) center symmetries.  The  breaking of the latter solves the problem.
In other words, it is the way  these anomalies (the impossibility of maintaining these symmetries in the UV) are matched by the system in the IR.

\subsection{Natural Anomaly Matching  \label{sec:NAM}}  

It was noted previously \cite{BKL2, BKL4, BKLDA} that, when dynamical symmetry breaking occurs,  the 
anomaly-matching is often satisfied entirely evidently, without the necessity of solving any arithmetic matching equations.
The precise way this  ``Natural Anomaly Matching''  works has been recently revisited and clarified in \cite{BKLO}. Let's review briefly  the argument here. 

Call the  set of  (left-handed) fermions of our model   collectively  as $\{\Psi\}$. They transform in some reducible representation of the color and flavor symmetry group   \footnote{As seen in all our examples, often the true internal group is not a product, as it contains discrete quotients.} $G_c \times G_f$.
At low energies  the  $G_c$  gauge interactions become strong.  
 Let us write the full set $\{\Psi\}$  as a direct sum
\be     \{\Psi\}  =    \{\Psi_0 \}    \oplus    \{\Psi_1\}  \;, 
\ee
where the first group of fermions  condense pairwise  
\be   \brc  {\Psi_0} {\Psi_0}  \ckt \sim   \Lambda^3  \;,  \label{condens} 
\ee
where $\Lambda$ is the renormalization-group invariant mass scale of the $G_c$ color gauge theory.  Due to the vacuum expectation values (VEV) (\ref{condens}) 
color and flavor symmetries are broken as 
\be     G_c \times  G_f  \to    {\tilde G}_c  \times     {\tilde G}_f\;. \label{LE}
\ee
In (\ref{LE}),  the unbroken gauge (${\tilde G}_c$)    and symmetry groups ${\tilde G}_f$  are such that 
\be      {\tilde G}_c \subset G_c\;, \qquad    {\tilde G}_f \subset   G_c/  {\tilde G}_c    \times  G_f\;.     \label{LE2}
\ee
Note that   the condensates (\ref{condens})  may break both (part of) color and flavor 
symmetry groups, but can leave some diagonal combination invariant, i.e., the color-flavor locking. 
Namely, the low-energy theory is a $ {\tilde G}_c$  
gauge theory with fermions $ \{\Psi_1\}$ in various representations of ${\tilde G}_f.$

If  the $ {\tilde G}_c$  theory is asymptotically free,  the system further evolves towards lower energies  (this is known as 
``tumbling'' \cite{Raby})  and the fate of the symmetries depend on the dynamics of  $ {\tilde G}_c$ interactions with fermions  $\{\Psi_1\}$. 
 The problem is essentially of the same nature as the original one:     we  assume from here on that 
$ {\tilde G}_c$   is either absent or infrared-free. 

Now pairs  of fermions participating in the condensates $ {\Psi_0} {\Psi_0}$    form together a singlet of  ${\tilde G}_f$, or in other words, 
they form together vector like representations of ${\tilde G}_f$. 
  They do not contribute to  ${\tilde G}_f^3 $    anomalies:
\be  {\Tr}\,   {\tilde T}^3|_{\Psi_0} =0\;.   \qquad     {\tilde T}  \subset   \{  T_c +  T_f \}\;. \label{enter} 
\ee
Next  note that for any generators ($T_f, T_c$)  of  $G_f$ and $G_c$,  a simple relation 
\be   \Tr    T_f^3  |_{{\Psi_0}+ {\Psi_1}}=  \Tr   (T_c+ T_f)^3  |_{{\Psi_0}+ {\Psi_1}}   \label{simple}   
\ee
holds.  To see this, recall that  the $G_c$  ($SU(N)$)  generators are traceless,  $G_f$ is ABJ-anomaly free, and the theory is free of gauge ($G_c$)   anomalies,  namely,
\be 
\Tr\left[T_f^2 T_c\right]|_{{\Psi_0}+ {\Psi_1}}  =  \Tr\left[T_c^2 T_f\right]|_{{\Psi_0}+ {\Psi_1}} =\Tr\left[T_c^3\right]|_{{\Psi_0}+ {\Psi_1}} =0\,. \ee
Now apply  (\ref{simple}) to   the generators $T_f$ and $T_c$  which enter the combination  $ {\tilde T} $  (see (\ref{enter})),  to get 
\be    { \Tr}\,    T_f^3  |_{{\Psi_0}+ {\Psi_1}}= { \Tr}\,     (T_c+ T_f)^3   |_{{\Psi_0}+ {\Psi_1}}=   { \Tr}\,   {\tilde T}^3 |_{{\Psi_0}+ {\Psi_1}} =  { \Tr}\,   {\tilde T}^3 |_{\Psi_1}  \;,    \label{above}
\ee
where  (\ref{enter}) has been used at the final step.

The relation ({\ref{above})  might look as the solution of the UV-IR anomaly matching condition.
Actually, this is not quite so. 
The problem is that at the scale $\Lambda$ the fermions $\{\Psi_1\}$ are strongly coupled, 
therefore in order to have the  ${\tilde G}_f$   symmetry properly realized at low energies, 
 all fermions in the set $\{\Psi_1\}$ must be replaced by  a set of $G_c$ invariant ``baryons''  $\{{\cal B}\} $ which are weakly coupled at
$\mu \lesssim  \Lambda$  and which reproduce all the  ${\tilde G}_f$  anomalies  due to $\{\Psi_1\}$, i.e., 
\be    {\Tr}\,   {\tilde T}^3|_{\Psi_1}   =   {\Tr}\,   {\tilde T}^3|_{\cal B}\;.  \label{theproblem}  
\ee
 But  this is the same as the original 't Hooft's
problem:  there is no  a priori guarantee that an appropriate set of baryons saturating  (\ref{theproblem})  exist or  can be found.

What happens in certain models with a color-flavor-locked dynamical Higgs phase \cite{ADS,BK,BKL2,BKL4, BKLZ2}  is that the VEV of the composite scalar 
$\Phi\sim  \Psi_0 \Psi_0$  acts  as a metric.  It can be used to convert the color to the flavor indices. In the simplest cases where 
\be    \brc  \Phi^i_a \ckt  =   \brc  {\Psi_0} {\Psi_0}  \ckt^i_a  \sim   \delta^i_a   \Lambda^3\;,   \label{CFlocking} 
\ee 
where  $i$ is the color index in the fundamental representation, ${\underline N}$,  of $G_c=SU(N)$, and $a$ is an index of the flavor 
symmetry $G_f$,   all color indices in  $\Psi_1$  can be eliminated (i.e., made gauge invariant) and  converted to flavor labels.  In other words, 
 all  $\Psi_1$ are replaced by  the ``baryons'' of the form,   
\be    \{{\cal B}\} \sim  \Psi \Phi\;, \quad \Psi \Phi^*\;, \quad       \Psi \Phi\Phi\;, \quad \  \Psi \Phi^*\Phi^* \;,\quad  \ldots\;,     \label{baryons}   
\ee
etc.  Which of the massless baryons among (\ref{baryons}) is the correct one, depends on the $G_c$ representation each of  $\Psi_1$ belongs to.  Once an appropriate set of the baryons are found, which replace in the infrared  all the UV fermions  $\Psi_1$,   
all the issues about perturbative and nonperturbative anomaly matching  disappear.

A second possible situation in which the problem (\ref{theproblem}) has a simple  solution is when the low-energy color gauge group
${\tilde G}_c$  
 in (\ref{LE}), (\ref{LE2}),    is infrared free.   This can occur if  an adjoint composite scalar field  takes a VEV  \footnote{This is similar to what happens in $\mathcal{N}=2$ supersymmetric gauge theories,  though in the latter it is an elementary adjoint scalar field that condenses.}. In this case, the condensate can either:
\begin{itemize}
    \item break completely the nonAbelian gauge symmetry to the Cartan subgroup, leaving an Abelian gauge theory in the IR     (dynamical Abelianization)  
    \be SU(N)_c \to U(1)^{N-1}\;,   \ee
    or  
    \item if there is enough ``active''  matter left in the infrared,  leave some IR-free but   nonAbelian gauge symmetry unbroken:
    \be
    SU(N)_c \to SU(N_1)_c \times SU(N_2) \times \ldots \times U(1)^{n}\;.
    \ee
\end{itemize} 
In both cases the elementary fermions in $\{\Psi_1\}$ survive into the infrared,  carrying all the information about the unbroken symmetry group ${\tilde G}_f$,  including the associated anomalies.  The anomaly matching  is  again fully   automatic.

\section{Dynamical symmetry breaking  \label{DSB}}  

The ideas exposed in Sec.~\ref{sec:implication} will now  be illustrated with a few examples in the context of the tensorial chiral gauge theories 
studied in  Sec.~\ref{Mixed}.

\subsection{Color-flavor locked dynamical Higgs phase  \label{CFlocking}}  

An example of color-flavor locking VEV in the $SU(5)$ tensor model and the consequent full solution of all anomaly matching problems,  was
recently given in  \cite{BKLO}.    This subsection is also borrowed from \cite{BKLO}.
 The matter fermions are   $\psi^{\{ij\}}$ and $\chi_{[ij]}^c$,   ($c=1,2,\ldots, 9$)  in symmetric and (anti-)  antisymmetric 2-index tensor representations,    
\begin{equation}\label{modelSU(5)}
\Yvcentermath1
\yng(2) \oplus    9  \; \bar{\yng(1,1)}\;. 
\end{equation}
The classical symmetry group is 
\begin{equation}\label{classical_flavor}
SU(5)_c \times  SU(9) \times U(1)_{\psi} \times U(1)_{\chi}\;,  
\end{equation}
with a
 non-anomalous combination of the two $U(1)$ groups being
\begin{equation}
U(1)_{\psi \chi}: \qquad  \psi \to e^{27i \alpha}\psi, \qquad   \chi \to e^{-7i\alpha}\chi\;,  \qquad   \alpha \in [0, 2\pi)    \label{Upc1} 
\end{equation}

The mixed anomalies of this model have been analyzed in Sec.~\ref{sec:su5}.    
As argued in Sec.~\ref{Mixed}, the most significant message from these anomalies is  that the  $U(1)_{\psi \chi}$ symmetry is spontaneously broken by some bifermion condensate. 
Let us assume, in view of the fact that  in this model, the most attractive channel is in the $\chi \chi$ composite in the fundamental representation, ${\underline 5}$,    that 
\be     \epsilon^{ijk \ell m}  \brc  \chi_{jk}^A     \chi_{\ell m}^9 \ckt    \propto  \delta^{i  A}  \ne 0\;,  \qquad A=1,2,\ldots, 5\;,  \qquad 
\brc \psi\chi \ckt =0\;. \label{condensBis} 
\ee
The symmetry  (\ref{classical_flavor})  is broken to  
\be      SU(5)_{cf} \times SU(3) \times  U(1) \times U(1)^{\prime}\;,   \label{Unbroken}  
\ee
where  $U(1) \times U(1)^{\prime}$ are two $U(1)$ symmetries unbroken by the  condensate, (\ref{condensBis}).

The color-flavor locking 
(\ref{condensBis}) means that the first 5 flavor indices, $A=1,2,\ldots, 5$  of  $\chi_{jk}^A$   act as the antifundamental
${\bar 5}$ of the color-flavor locked $SU(5)$.  
     Let us indicate the bifermion composite scalar  (\ref{condensBis})   as 
\be (\chi \chi)^i_A   \sim     \epsilon^{ijk \ell m}  \chi_{jk}^A     \chi_{\ell m}^9 \;.\ee
The massless baryons we expect in the infrared are 
\be    B^{\{A_1 A_2\}} =   \psi^{\{ij\}}  \overline {(\chi \chi)^i_{A_1} }  \overline {(\chi \chi)^j_{A_2} } \;;
\ee
\be 
 {\hat B}_{[A_1 A_2], A} =     \chi_{[i  j]}^A   \,   (\chi \chi)^i_{A_1}   (\chi \chi)^j_{A_2}\;, 
\ee
(where the indices  $A, A_1, A_2$ run up to $5$,   forming  \be \Yvcentermath1  {\bar  { \yng(2,1)}}  \ee  of the flavor $SU(5)\subset SU(9)$);   
and 
\be {\hat B}_{[A_1 A_2], B} =     \chi_{[i  j]}^B   \,   (\chi \chi)^i_{A_1}   (\chi \chi)^j_{A_2}\;, 
\ee
where  the indices  $A_1, A_2$ run up to $5$,  $B=6,7,8$.

Table~\ref{decomp19}  gives a complete solution of anomaly matching problem.
Note that,  a priori,    finding the solution of the anomaly matching requirement with respect to the unbroken symmetry group   (\ref{Unbroken}), including its global structures, not explicitly shown there,  with the known UV fermions  (the upper half of  Table~\ref{decomp19}),  
would have been a formidable task.

  \begin{table}[h!t]
  \centering 
  \begin{tabular}{|c|c|c |c|c|c|  }
\hline
$ \phantom{{{   {  {\yng(1)}}}}}\!  \! \! \! \! \!\!\!$   & fields   &  $SU(5)_{\rm cf} $      &  $SU(3)$   &  $ U(1)   $      &   $ U(1)^{\prime}$  \\
 \hline
   \phantom{\huge i}$ \! \!\!\!\!$  {\rm UV} &  $\psi^{\{i  j\}} $    &     $ \Yvcentermath1 { { \yng(2)}} $     &   $ 15 \cdot (\cdot)$    & $0$    &   $3$   \\[1.5ex]
  & $\chi_{[i  j]}^A$   &   $\Yvcentermath1 {\bar  { \yng(2,1)}}  \oplus  {\bar  { \yng(1,1,1)}} $    &   $ 50  \cdot  (\cdot) $    & $  -3$   &  $-7$   \\[1.5ex]
  & $\chi_{[i  j]}^B$   &   $ 3 \cdot  \Yvcentermath1 {\bar  { \yng(1,1)}} $    &   $  10  \cdot   \Yvcentermath1 {\yng(1)} $    & $  4$   &  $7$   \\[1.5ex]
  & $\chi_{[i  j]}^9$   &   $\Yvcentermath1  {\bar  { \yng(1,1)}} $    &   $  10  \cdot  (\cdot) $    & $  3$   &   $7$    \\[1.5ex]
    \hline 
 $ \phantom{{\bar{ \bar  {\bar  {\yng(1,1)}}}}}\!  \! \!\! \! \!  \!\!\!$  {\rm IR}  &   $ B^{\{A_1 A_2\}} $      &   $\Yvcentermath1 { { \yng(2)}} $     &    $    15 \cdot  (\cdot) $       &    $ 0 $    &   $3$     \\[1.5ex]
  &    ${\hat B}_{[A_1 A_2], A}$        &   $ \Yvcentermath1 {\bar  { \yng(2,1)}}  $     &    $   40  \cdot  (\cdot) $       &    $ -3 $    &     $-7$   \\[1.5ex]
 &    ${\hat B}_{[A_1 A_2], B}$      &    $ 3 \cdot \Yvcentermath1 {\bar  {\yng(1,1)}} $      &    $  10  \cdot \Yvcentermath1 {\yng(1)}  $    &    $4$  &    $7$  \\[1.5ex]
\hline
\end{tabular}  
  \caption{\footnotesize  CF locking (\ref{condensBis}) and  the natural anomaly matching  in the model  (\ref{modelSU(5)}).   
  The index $A$  (and $A_{1,2}$) run up to $5$;     the index $B$ runs over  $B=6,7,8$. 
  The totally antisymmetric part of  $\chi_{[i  j]}^A$  and $\chi_{[i  j]}^9$  pair up to condense,  form a massive Dirac fermion, and decouple.  
  The rest of the UV fermions are completely matched by the three types of baryons in the IR, in their quantum numbers and multiplicities.   Any conventional
  as well as  generalized anomaly matching is automatic. 
  }
   \label{decomp19}
\end{table}

\subsection{Dynamical Abelianization  \label{dynAb}}

Another possible phase is dynamical Abelianization, analogous to what is believed to take place in the  $\psi\chi\eta$  model \cite{BKLDA}. 
Let us consider the  $SU(6)$ model   (\ref{su6model})    studied in Sec.~\ref{sec:su6}.  The formation of a bifermion condensate $\psi \chi$ in the adjoint representation of $SU(6)$, which is the favored one according to the MAC (the maximally-attractive-channel)  criterion \cite{Raby} in this model, is assumed. The condensate breaks the gauge group as follow:
\begin{equation}\label{dyn_ab_SU(6)}
SU(6) \rightarrow U(1)^5 = U(1)_1 \times U(1)_2 \times U(1)_3 \times U(1)_4 \times U(1)_5
\end{equation}
The generators of the Cartan sub-algebra are:

\begin{equation}\label{Cartan_sub_SU(6)}
\begin{split}
& t^1  = \frac{1}{2} \begin{bmatrix}
								1 & & & & & \\
								& -1 & & & & \\
								& & 0 & & & \\
								& & & 0 & & \\
								& & & & 0 & \\
								& & & & & 0
\end{bmatrix}\, , \quad 
t^2  = \frac{1}{2 \sqrt{3}}\begin{bmatrix}
								1 & & & & & \\
								& 1 & & & & \\
								& & -2 & & & \\
								& & & 0 & & \\
								& & & & 0 & \\
								& & & & & 0
\end{bmatrix} \, , \quad
t^3 =\frac{1}{2 \sqrt{6}} \begin{bmatrix}
								1 & & & & & \\
								& 1 & & & & \\
								& & 1 & & & \\
								& & & -3 & & \\
								& & & & 0 & \\
								& & & & & 0
\end{bmatrix} \, ,  \\
& t^4 =\frac{1}{2 \sqrt{10}} \begin{bmatrix}
								1 & & & & & \\
								& 1 & & & & \\
								& & 1 & & & \\
								& & & 1 & & \\
								& & & & -4 & \\
								& & & & & 0
\end{bmatrix} \, , \quad
t^5 =\frac{1}{2 \sqrt{15}} \begin{bmatrix}
								1 & & & & & \\
								& 1 & & & & \\
								& & 1 & & & \\
								& & & 1 & & \\
								& & & & 1 & \\
								& & & & & -5
\end{bmatrix} \, .
\end{split}
\end{equation}
The bifermion condensate also breaks the flavor symmetry as:
\begin{equation}\label{flav_break_SU(6)}
SU(5) \times U(1)_{\psi \chi} \rightarrow SU(4) \times U(1)' \, ,
\end{equation}
where $U(1)'$ is the sub-group of $U(1)_{\psi \chi} \times U(1)_d$ left unbroken by the condensate. $U(1)_d \subset SU(5)$ is generated by
\begin{equation}\label{gen_U(1)_d_SU(6)}
\begin{bmatrix}
								\mathbbm{1}_{4\times 4} & \\
								& -4 
\end{bmatrix} \, .
\end{equation}
In the IR, the matter content reorganizes as  in Table \ref{table_dyn_ab_SU(6)}.

\begin{table}[H]
\begin{center}
\begin{tabular}{cccccccc}
\toprule
fields & $U(1)_1$ & $U(1)_2$ & $U(1)_3$ & $U(1)_4$ & $U(1)_5$ & $SU(4)$ & $U(1)'$\\
\midrule
$\psi^{11}$ & 1 & $1/\sqrt{3}$ & $1/\sqrt{6}$ & $1/\sqrt{10}$ & $1/\sqrt{15}$ & $(\cdot)$ & 5\\[1.5ex]
$\psi^{22}$ & $-1$ & $1/\sqrt{3}$ & $1/\sqrt{6}$ & $1/\sqrt{10}$ & $1/\sqrt{15}$ & $(\cdot)$ & 5\\[1.5ex]
$\psi^{33}$ & 0 & $-2/\sqrt{3}$ & $1/\sqrt{6}$ & $1/\sqrt{10}$ & $1/\sqrt{15}$ & $(\cdot)$ & 5\\[1.5ex]
$\psi^{44}$ & 0 & 0 & $-3/\sqrt{6}$ & $1/\sqrt{10}$ & $1/\sqrt{15}$ & $(\cdot)$ & 5\\[1.5ex]
$\psi^{55}$ & 0 & 0 & 0 & $-4/\sqrt{10}$ & $1/\sqrt{15}$ & $(\cdot)$ & 5\\[1.5ex]
$\psi^{66}$ & 0 & 0 & 0 & 0 & $-5/\sqrt{15}$ & $(\cdot)$ & 5\\
\midrule
$\chi^c_{12}$ & 0 & $-1/\sqrt{3}$ & $-1/\sqrt{6}$ & $-1/\sqrt{10}$ & $-1/\sqrt{15}$ & \Yvcentermath1 $\yng(1)$ & $-5/4$\\[1.5ex]
$\chi^c_{13}$ & $-1/2$ & $1/2\sqrt{3}$ & $-1/\sqrt{6}$ & $-1/\sqrt{10}$ & $-1/\sqrt{15}$ & \Yvcentermath1 $\yng(1)$ & $-5/4$\\[1.5ex]
$\chi^c_{14}$ & $-1/2$ & $-1/2\sqrt{3}$ & $1/\sqrt{6}$ & $-1/\sqrt{10}$ & $-1/\sqrt{15}$ & \Yvcentermath1 $\yng(1)$ & $-5/4$\\[1.5ex]
$\chi^c_{15}$ & $-1/2$ & $-1/2\sqrt{3}$ & $-1/2\sqrt{6}$ & $3/2\sqrt{10}$ & $-1/\sqrt{15}$ & \Yvcentermath1 $\yng(1)$ & $-5/4$\\[1.5ex]
$\chi^c_{16}$ & $-1/2$ & $-1/2\sqrt{3}$ & $-1/2\sqrt{6}$ & $-1/2\sqrt{10}$ & $2/\sqrt{15}$ & \Yvcentermath1 $\yng(1)$ & $-5/4$\\[1.5ex]
$\chi^c_{23}$ & $1/2$ & $1/2\sqrt{3}$ & $-1/\sqrt{6}$ & $-1/\sqrt{10}$ & $-1/\sqrt{15}$ & \Yvcentermath1 $\yng(1)$ & $-5/4$\\[1.5ex]
$\chi^c_{24}$ & $1/2$ & $-1/2\sqrt{3}$ & $1/\sqrt{6}$ & $-1/\sqrt{10}$ & $-1/\sqrt{15}$ & \Yvcentermath1 $\yng(1)$ & $-5/4$\\[1.5ex]
$\chi^c_{25}$ & $1/2$ & $-1/2\sqrt{3}$ & $-1/2\sqrt{6}$ & $3/2\sqrt{10}$ & $-1/\sqrt{15}$ & \Yvcentermath1 $\yng(1)$ & $-5/4$\\[1.5ex]
$\chi^c_{26}$ & $1/2$ & $-1/2\sqrt{3}$ & $-1/2\sqrt{6}$ & $-1/2\sqrt{10}$ & $2/\sqrt{15}$ & \Yvcentermath1 $\yng(1)$ & $-5/4$\\[1.5ex]
$\chi^c_{34}$ & 0 & $1/\sqrt{3}$ & $1/\sqrt{6}$ & $-1/\sqrt{10}$ & $-1/\sqrt{15}$ & \Yvcentermath1 $\yng(1)$ & $-5/4$\\[1.5ex]
$\chi^c_{35}$ & 0 & $1/\sqrt{3}$ & $-1/2\sqrt{6}$ & $3/2\sqrt{10}$ & $-1/\sqrt{15}$ & \Yvcentermath1 $\yng(1)$ & $-5/4$\\[1.5ex]
$\chi^c_{36}$ & 0 & $1/\sqrt{3}$ & $-1/2\sqrt{6}$ & $-1/2\sqrt{10}$ & $2/\sqrt{15}$ & \Yvcentermath1 $\yng(1)$ & $-5/4$\\[1.5ex]
$\chi^c_{45}$ & 0 & 0 & $3/2\sqrt{6}$ & $3/2\sqrt{10}$ & $-1/\sqrt{15}$ & \Yvcentermath1 $\yng(1)$ & $-5/4$\\[1.5ex]
$\chi^c_{46}$ & 0 & 0 & $3/2\sqrt{6}$ & $-1/2\sqrt{10}$ & $2/\sqrt{15}$ & \Yvcentermath1 $\yng(1)$ & $-5/4$\\[1.5ex]
$\chi^c_{56}$ & 0 & 0 & 0 & $2/\sqrt{10}$ & $2/\sqrt{15}$ & \Yvcentermath1 $\yng(1)$ & $-5/4$\\
\bottomrule
\end{tabular}
\caption{\footnotesize  Representations and charges of the matter content of the IR theory with respect  to the unbroken gauge and flavor symmetries. }\label{table_dyn_ab_SU(6)}
\end{center}
\end{table}

\subsection{Low-energy nonAbelian gauge group    \label{sec:nonAbelian}} 

More generally,  it is possible that the strong gauge group is dynamically broken by bifermion condensates
$\ev{\psi\chi}$  of a diagonal form in the adjoint representation,  but with the symmetry breaking pattern,
\beq
SU(N) \to SU(n) \times \dots   \label{generalDSB}
\eeq
The $SU(n)$ factor  in (\ref{generalDSB}) may  remain infrared-free  (weakly coupled) at low energies, depending on the model.   By decomposing the fermions  (\ref{models})  as direct sums of the irreducible representations  of  the $SU(n)$ subgroup,  
the (first coefficient the)  beta function of $SU(n)$ can be seen to be \cite{BKLproc, corfu} 
\bea
\beta(SU(n)) &=& 11 \cdot n - \frac{N-4}{\ell}\cdot (n+2) - \frac{N+4}{\ell}\cdot (n-2) - \frac{8}{\ell}\cdot(N-n)\cdot 1   \nonumber \\  &=& \frac{16+8n+11 \ell n -8 N -2 n N}{\ell}\;.  
\eea
The sign change occurs at 
\beq
n^*= \frac{8N -16}{8 + 11\ell -2N}
\eeq
so that    the integer part of $n^*$, $[n^*]$ is the largest $n$ that can be  IR free. Clearly if $[n^*]=1$ there are no non-Abelian IR free (gauge) symmetry breaking patterns possible. In the models studied in Sec.~\ref{Mixed}: 

\begin{itemize}
\item  For the $SU(5)$  model  of Sec.~\ref{sec:su5},    $N=5, \,  \ell=1$,   
$  n^*=   \frac{8}{3}$,    so $[n^*]=2$. Indeed 
\beq
\beta(SU(2)) =-6\;.
\eeq
The possible nonAbelian, IR free gauge symmetry breaking is 
\beq
 SU(5)   \to   SU(2)\times SU(2) \times  U(1)^2\;.   \label{IRfree}  
\eeq
\item  In the $SU(6)$  model  of Sec.~\ref{sec:su6},     
   $N=6, \,\ell=2$,      $n^*=    \frac{16}{9}$.    As  $[n^*]=1$  there are no non-Abelian IR free symmetry breaking patterns
  possible.    Abelianization is the only IR-free possibility.

  \item   In the    ($N=8$, $\ell=4$)   $SU(8)$  model  of Sec.~\ref{sec:su8},      $n^*=    \frac{4}{3}$.  Again,  no symmetry breaking into 
  nonAbelian low-energy system is possible.

   \end{itemize}

With a larger number of fermions in the model,    infrared nonAbelian gauge groups appear to be certainly a dynamical possibility: 

	\begin{itemize}
   \item $N=6$,    $\ell=1$,   $n^*=    \frac{32}{7}
$,    so $[n^*]=4$, and 
\beq
\beta(SU(4)) =-4\;. 
\eeq
A possible IR free breaking mode  is 
\beq
 SU(6)   \to   SU(4)\times SU(2) \times  U(1)  \;.
\label{b16}
\eeq

\item $N=8$,  $\ell=2$,   $n^*=  3+ \frac{3}{7}$   and  
\beq
\beta(SU(3)) =-3\;.  
\eeq
A possible IR free breaking pattern is
\beq
 SU(8)   \to   SU(3)\times SU(3)\times SU(2) \times  U(1)^2\;,
\label{b18}
\eeq
and so on. 

	   \end{itemize}

In the case of the $SU(5)$ model,  (\ref{IRfree}),   the fermions in the IR which take care of all of the anomalies of the 
	unbroken gauge and flavor symmetry  group, are shown in Table~\ref{table_IR_SU(5)}, taken from   
	\cite{BKLO}.
\begin{table}[H]
\begin{center}
\begin{tabular}{ccccccc}
\toprule
fields & $SU(2)_1$ & $SU(2)_2$ & $U(1)_1$ & $U(1)_2$ & $SU(8)$ & $U(1)'$\\
\midrule
$\psi^{\lbrace ij \rbrace}$ & \Yvcentermath1 $\yng(2)$ & $3 \hspace{0.5ex} (\cdot)$ & 4 & 2 & $3 \hspace{0.5ex} (\cdot)$ & 27\\
[1.5ex]
$\psi^{\lbrace IJ \rbrace}$ & $3 \hspace{0.5ex} (\cdot)$ & \Yvcentermath1 $\yng(2)$ & $-2$ & $-8$ & $3 \hspace{0.5ex} (\cdot)$ & 27\\
[1.5ex]
$\psi^{55}$ & $(\cdot)$ & $(\cdot)$ & $-4$ & 12 & $(\cdot)$ & 27\\
\midrule
$\chi^9_{\left[ ij\right] }$ & $(\cdot)$ & $(\cdot)$ & $-4$ & $-2$ & $(\cdot)$ & $-27$\\
[1.5ex]
$\chi^9_{\left[ IJ \right] }$ & $(\cdot)$ & $(\cdot)$ & 2 & 8 & $(\cdot)$ & $-27$\\
[1.5ex]
\midrule
$\chi^c_{\left[ ij\right] }$ & $8 \hspace{0.5ex} (\cdot)$ & $8 \hspace{0.5ex} (\cdot)$ & $-4$ & $-2$ & \Yvcentermath1 $\yng(1)$ & $-9/2$\\
[1.5ex]
$\chi^c_{iJ}$ & \Yvcentermath1 $16 \hspace{0.5ex} \yng(1)$ & \Yvcentermath1 $16 \hspace{0.5ex} \yng(1)$ & $-1$ & 3 & \Yvcentermath1 $4 \hspace{0.5ex} \yng(1)$ & $-9/2$\\
[1.5ex]
$\chi^c_{\left[ IJ \right] }$ & $8 \hspace{0.5ex} (\cdot)$ & $8 \hspace{0.5ex} (\cdot)$ & 2 & 8 & \Yvcentermath1 $\yng(1)$ & $-9/2$\\
[1.5ex]
$\chi^c_{i5}$ & \Yvcentermath1 $8 \hspace{0.5ex} \yng(1)$ & $16 \hspace{0.5ex} (\cdot)$ & 0 & $-7$ & \Yvcentermath1 $2 \hspace{0.5ex} \yng(1)$ & $-9/2$\\
[1.5ex]
$\chi^c_{I5}$ & $16 \hspace{0.5ex} (\cdot)$ & \Yvcentermath1 $8 \hspace{0.5ex} \yng(1)$ & 3 & $-2$ & \Yvcentermath1 $2 \hspace{0.5ex} \yng(1)$ & $-9/2$\\
\bottomrule
\end{tabular}
\caption{\footnotesize  Representations and charges of the matter content in the IR theory in respect of the continuous gauge and flavor symmetries unbroken by the condensate. The color indices run as  $i,j=1,2$,   $I,J=3,4$.    The multiplicities are also indicated. The fields $\psi^{iJ}$-$\chi^9_{iJ}$, $\psi^{i5}$-$\chi^9_{i5}$ and $\psi^{I5}$-$\chi^9_{I5}$ form massive Dirac pairs and decoupled. 
}\label{table_IR_SU(5)}
\end{center}
\end{table}

\subsection{The  $a$-theorem, the ACS and MAC criteria}

  Arguments based solely on the symmetry considerations such as ours, are unfortunately not strong enough to allow us to pinpoint a particular phase as the correct one realized by the system dynamically in the infrared.  In such a situation it could be of some benefit to 
keep in mind  several qualitative dynamical criteria (and constraints) proposed in the past,  to see whether or not a dynamical scenario being considered  is at least plausible, or in some case, is to be excluded.

Let us  briefly review here, taking the   $(N, \ell) = (5, 1)$ model of Sec.~\ref{sec:su5} as an illustrative example,  
 various coefficients appearing in the  
  the $a$-theorem \cite{Cardy:1988cwa, Komargodski:2011vj} and     in    the   
  Appelquist-Cohen-Schmaltz  (ACS)  prescription (the free-energy $f$  in  the UV and in the IR) \cite{ACS, ACSS,ADS},    as well as  in the MAC  (the most attractive channel)   criterion 
 \cite{Raby}.   
$a$  and  $f$, given by 
\begin{equation}\label{a-theorem}
a = \frac{1}{360} \left[ \left( \symbol{35} \, scalar \, bosons \right) + \frac{11}{2} \left( \symbol{35} \, Weyl \, fermions \right) + 62 \left( \symbol{35} \, vector \, bosons \right) \right] 
\end{equation}
and
\begin{equation}\label{ACS_pres}
f = \left( \symbol{35} \, bosons \right) + \frac{7}{4} \left( \symbol{35} \, Weyl \, fermions \right) \, ,
\end{equation}
take the values reported in  Table~\ref{table_coeff}.    The $a$-theorem states that $a_{IR} \leq a_{UV}$, whereas  the ACS criterion demands  that $f_{IR} \leq f_{UV}$.   All possible phases  
considered are seen to  satisfy these criteria. 
\begin{table}[H]
\begin{center}
\begin{tabular}{cc|ccc}
\toprule
& UV       & CF-locked phase & Non-Ab. Gauge group & Dyn. Abelianization \\
\midrule
$\tilde{a}$ & 2065.5 & 514.5  & 1002.5 & 732.5 \\
$f$ & 231.75 & 195.75 & 188.75 & 173.75 \\
\bottomrule
\end{tabular}
\caption{ \footnotesize   Values of $\tilde{a}$ and $f$ for the UV $(N, \ell) = (5, 1)$ model and its IR phases studied in this work.
 The scaled coefficient $\tilde{a}$   is defined by $\tilde{a} \equiv 360 a$. 
}\label{table_coeff}
\end{center}
\end{table}

In Table \ref{table_MAC},  the MAC   criterion, i.e., to maximizing the
absolute value of   $\Delta C_2 \equiv C_2(\mathcal{R}_c) - C_2(\mathcal{R}_1) - C_2(\mathcal{R}_2)$,     is studied for the different candidate  phases.  $C_2(\mathcal{R})$ is the quadratic Casimir of the representation $\mathcal{R}$.

\begin{table}[H]
\begin{center}
\begin{tabular}{cccc}
\toprule
Phase  &  Condensate    & $\mathcal{R}_c$ &  $\Delta C_2 \vert_{N=5}$\\
\midrule
  CF-locked (dyn Higgs) phase    &  $\langle \chi \chi \rangle   $    & $\Yvcentermath1 \yng(1)$ &  $-4.8$\\[1.5ex]
 Abelian or nonAbelian IR free phase     &  $\langle  \psi \chi \rangle $    & adjoint &  $-4.2$\\[1.5ex]
\bottomrule
\end{tabular}
\caption{ \footnotesize    Condensates and the MAC criterion in different phases discussed in this work:  
CF-locked dynamical Higgs phase (Sec.~\ref{CFlocking}),   Dynamical Abelianization (analogue of  Sec.~\ref{dynAb}),  
and the infrared nonAbelian gauge group (Sec.~\ref{sec:nonAbelian}).  
 $\mathcal{R}_c$ is the representation to which the bifermion composite belongs.
 }\label{table_MAC}
\end{center}
\end{table}

In conclusion,  the MAC criterion would suggest that the system in the infrared is  in  the  color-flavor locked Higgs  phase studied in Sec.~\ref{CFlocking}.
The free energy $f$  in the infrared  is found to be the  largest  among the different possible phases, whereas  ${\tilde a}$   the smallest, meaning that the RG flow is the steepest.

\section{Conclusion}

The dynamics of tensorial chiral gauge theories  (\ref{models})  has been discussed. Even though the analysis  of the mixed anomalies by taking into account the presence of color-flavor locked center symmetries,  is itself  a straightforward extension of the studies made for  other types of chiral gauge theories \cite{BKL1}-\cite{corfu},  it turns out to be a rather subtle problem to draw the correct conclusion from the  results of the anomaly calculations. What  {\it   is}  the implication of the mixed anomalies found?  

 The main section of this work, Sec.~\ref{sec:implication}, discussed this question.  We argue that the relevant question is what the various mixed anomalies found imply about the most likely dynamical effects of the strong gauge interactions.  Confinement or (dynamical) Higgs phase, dynamical Abelianization (Coulomb phase),  spontaneous flavor symmetry breaking,  bi-fermion or multi-fermion condensate formation?    
  Compared to this central question, the details of the mixed anomalies  involving various discrete symmetries of a given model, turn out to be a secondary, and even irrelevant, one.   

In our view, the main implication of the mixed anomalies in these tensorial chiral gauge theories  is that the (nonanomalous) $U_{\psi\chi}(1)$ symmetry is spontaneously broken in the infrared, by some bifermion condensates. Such condensates break dynamically (part or all of) the color gauge symmetry.  And this, in turn, means that the color-flavor locked $1$-form center symmetry  (${\mathbbm Z}_N$) itself is also broken by the vacuum.   At this point,  
		\begin{description}
  \item[(a)]  Anomaly matching with respect to the symmetries which are not spontaneously broken, is  
  automatic, due to the Natural Anomaly Matching mechanism \cite{BKLO}, reviewed in Sec.~\ref{sec:NAM}.  This includes not only the conventional 't Hooft anomalies, but also the new, generalized anomalies, as well as global anomalies;   
  \item[(b)]   The details of the mixed anomalies involving various discrete symmetries and 1-form  center symmetry, are immaterial.  
  The breaking of the color-flavor locked $1$-form center symmetry  (${\mathbbm Z}_N$) by the vacuum is the way the system ``matches''
  those mixed anomalies found in the UV  in the IR,  but otherwise  no further constraints on the detailed symmetry properties of the IR effective theory are
  implied.

\end{description}

	\section*{Acknowledgment} 
	This work is supported by the  INFN special initiative grant, GAST (Gauge and String Theories).

%

\end{document}